\pdfoutput=1
\documentclass[cernpreprint, atlasdraft=false, texlive=2023, UKenglish, texmf, orcidlogo]{atlasdoc}
 
\usepackage[subcaption=true]{atlaspackage}
\usepackage{atlasbiblatex}
 
\usepackage{atlasphysics}

\usepackage{verbatim}
 
\graphicspath{{logos/}{figures/}}
 
\usepackage{ANA-EXOT-2019-12-PAPER-defs}

\AtlasTitle{Search for pair-produced scalar and vector leptoquarks decaying into third-generation quarks and first- or second-generation leptons in $pp$ collisions
with the ATLAS detector
}
 
\AtlasAbstract{
A search for pair-produced scalar and vector leptoquarks decaying into quarks and leptons of different generations is presented. It uses
the full LHC Run~2\xspace (2015--2018) data set of \SI{139}{\mbox{fb\(^{-1}\)}} collected with the ATLAS detector in proton--proton collisions
at a centre-of-mass energy of $\sqrt{s}=\SI{13}{\text{Te\kern-0.1em V}}$.
Scalar leptoquarks with charge $-(1/3)e$ as well as scalar and vector leptoquarks with charge $+(2/3)e$ are considered.
All possible decays of the pair-produced leptoquarks into quarks of the third generation ($t, b$) and charged or neutral leptons of the first or
second generation ($e, \mu, \nu$) with exactly one electron or muon in the final state are investigated.
No significant deviations from the Standard Model expectation are observed. Upper limits on the production cross-section
are provided for eight models as a function of the leptoquark mass and the branching ratio of the leptoquark into the charged or neutral lepton.
In addition, lower limits on the leptoquark masses are derived for all models across a range of branching ratios.
Two of these models have the goal of providing an explanation for the recent $B$-anomalies.
In both models,
a vector leptoquark decays into charged
and neutral leptons of the second generation with a similar branching fraction.
Lower limits of 1980~GeV and 1710~GeV are set on the leptoquark mass for these two models.
}
 
\author{The ATLAS Collaboration}
 
\AtlasRefCode{EXOT-2019-12}

\AtlasJournal{JHEP}
\AtlasJournalRef{JHEP 06 (2023) 188}
\AtlasDOI{10.1007/JHEP06(2023)188}

\PreprintIdNumber{CERN-EP-2022-145}

\hypersetup{pdftitle={ATLAS document},pdfauthor={The ATLAS Collaboration}}
 
\begin{document}
 
\maketitle
 
\tableofcontents
 
\section{Introduction}


Leptoquarks (LQs) have already been discussed for a few decades, as they provide a connection between the quark and lepton sectors, which exhibit similar structures.
They are predicted by many extensions
of the Standard Model (SM), e.g.\ in unified theories \cite{georgi_glashow_unification,pati_salam_colour, SU5_lq_constr} and technicolor \cite{dimop_suss,techni2,techni3} or composite models \cite{composite_models_initial_1, composite_models_initial_2, SCHREMPP1985101}.
Recent hints of a potential violation of lepton flavour universality  in various measurements of $B$-meson decays (`\banomalies')
\cite{babar,belle,lhcb1,Aaij:2015oid,Hirose:2016wfn,Belle:2016ure,Wehle:2016yoi,lhcb2,LHCb:2017smo,Aaij:2017tyk,LHCb:2019hip,Aaij:2021vac}
can also be attributed, if confirmed,
to the exchange
of leptoquarks~\cite{Hiller:2014yaa, Gripaios:2014tna, Freytsis:2015qca, DiLuzio:2017chi, Buttazzo:2017ixm, Cline:2017aed, Bauer:2015knc, PhysRevD.95.055018}\footnote{After submission of the paper a new result from LHCb has been reported~\cite{LHCb:2022qnv, LHCb:2022zom} superseding Refs.~\cite{lhcb2,LHCb:2019hip,Aaij:2021vac} and indicating compatibility with the SM.}. In addition, some of these models~\cite{Bauer:2015knc, PhysRevD.95.055018,Chakraverty_2001} introducing LQs aim
to simultaneously provide an explanation for the longstanding discrepancy between the measured and the predicted anomalous magnetic moment of the muon \cite{PhysRevLett.126.141801}.
LQs are bosons carrying colour charge and fractional electrical charge. They possess non-zero baryon and lepton numbers and are assumed to decay into a quark--lepton pair.
The branching ratio into a quark and a charged lepton is denoted by \br, and that into a quark and neutrino by $1-\br$.
Leptoquarks can be scalar or vector bosons and can be produced singly or in pairs in proton--proton collisions.

The assumption that LQs can only interact with leptons and quarks of the same generation   
\cite{Buchmuller:1986zs}, and are spin-0 particles, has been used in most of the searches for LQs.
Recently, however, searches for LQs with couplings to quarks and leptons of different generations have aroused interest
because these couplings are required in order to explain the \banomalies
mentioned above.
The theoretical explanations usually require LQs with couplings to third-generation quarks and second-generation leptons.
In particular, vector LQs with a charge of 2/3, in units of the elementary charge $e$, have been identified recently as promising
candidates~\cite{Alonso:2015sja,Barbieri:2015yvd}.

First results of searches for LQs with couplings to quarks and leptons of different generations were published
recently for pair-produced LQs decaying into charged leptons, i.e.\ in dilepton final states.
Results from the full LHC \RunTwo data set (2015--2018) of proton--proton collisions at $\sqrt{s}=\SI{13}{\TeV}$, corresponding to an integrated luminosity of \SI{139}{\ifb}, are available for pair-produced
scalar LQs decaying into a \topquark and an electron or muon from the ATLAS~\cite{EXOT-2019-19} and the CMS Collaborations~\cite{CMS-EXO-21-002} and for the decay into a $b$-, $c$- or light-quark and an electron or muon~\cite{EXOT-2019-13} from ATLAS.
In addition, using a partial (\SI{35.9}{\ifb}) \RunTwo data set, CMS has published a result for pair-produced vector LQs decaying into a \topquark and a muon, obtained by scaling the results for scalar LQs to the larger production cross-sections expected for vector LQs, assuming no kinematic differences~\cite{CMS-B2G-16-027}.
 
\begin{figure}
\begin{center}
\begin{subfigure}[b]{0.5\textwidth}
\centering
\includegraphics[width=0.9\linewidth]{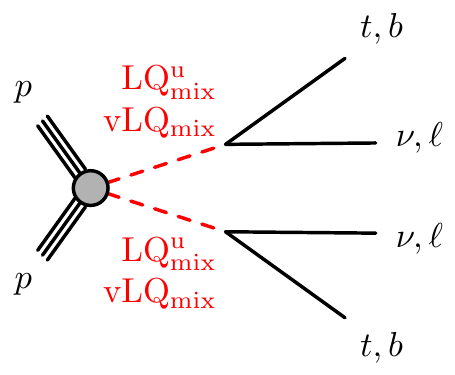}
\caption{}
\label{fig:uptypeLQ}
\end{subfigure}\begin{subfigure}[b]{0.5\textwidth}
\centering
\includegraphics[width=0.9\linewidth]{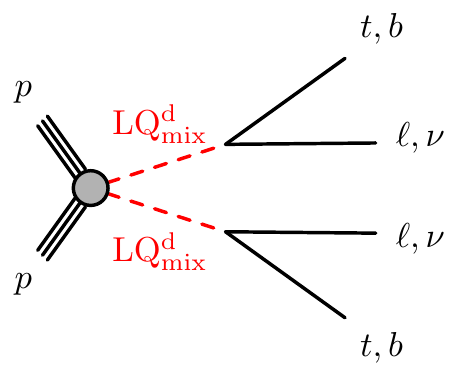}
\caption{}
\label{fig:downtypeLQ}
\end{subfigure}
\caption{
Pair production and decay of \subref{fig:uptypeLQ} up-type scalar (\lqmixu) and vector (\vLQ) LQs and \subref{fig:downtypeLQ} down-type scalar (\lqmixd) LQs with
$\ell = e, \mu$. No distinction is made between particles and antiparticles.
}
\label{fig:feynman-diagrams}
\end{center}
\end{figure}

Given the good coverage of large branching ratios \br by the dileptonic measurements described above, the results presented
here use a single-lepton (electron or muon) final state optimised for medium to small \br. In this case, one of the LQs decays
into a neutrino and the other decays into a charged lepton,
or both decay into neutrinos and the charged lepton arises from a leptonically decaying \topquark ($t_{\text{lep}}$).
The results are interpreted as searches for pair-produced LQs with charges of either $\pm(2/3)e$ (up-type) or $\pm(1/3)e$ (down-type).
All possible decays of the pair-produced up-type
and down-type
LQs into a quark ($t$, $b$) of the third generation and a lepton ($\ell$, $\nu$) of the first or second generation are considered, as seen in~\Cref{fig:feynman-diagrams}.
With flavour off-diagonal couplings allowed, the model used for up-type
(\lqmixu) and down-type (\lqmixd) scalar LQs
is an extension
of that~\cite{Mandal:2015lca} used in
previous ATLAS searches~\cite{EXOT-2017-30},
where all possible decays of the pair-produced up-type and down-type
scalar LQs into a quark ($t$, $b$) and a lepton ($\tau$, $\nu$) of the third generation were considered.
The present search for up-type LQs is also optimised
for a vector LQ (vLQ) model~\cite{Baker2019} designed to provide an explanation for the various \banomalies.

The analysis strategy is based on a final-state signature with one lepton, high
missing transverse momentum and at least four jets due to a hadronically decaying top-quark ($t_{\text{had}}$) and a \bquark.
Dedicated neural networks (NNs), trained in a common training region, are used for the separation of signal and background.
This is done separately for scalar and vector LQ pair-production because they exhibit different kinematic behaviour
for small values of \br, i.e.\ when the charged lepton arises mostly from the \topquark decay.
For each of the models and for various branching ratios, a signal region (SR) based on the NN output is defined.
Control regions (CRs) are defined so as to be enriched in the various background processes.
They are orthogonal to the
SR, and orthogonal to each other. The statistical interpretation
is based on a simultaneous fit to the CRs and the SR, in which the normalisations for top-quark pair (\ttbar), \Wjets, and single \topquark production as well as a possible signal contribution are determined, while taking into account the experimental and theoretical systematic uncertainties.
The results are presented as limits on the leptoquark mass as a function of the branching ratio.


\section{ATLAS detector}

The ATLAS experiment~\cite{PERF-2007-01} at the LHC is a multipurpose particle detector with a forward--backward symmetric cylindrical geometry and a near \(4\pi\) coverage in solid angle.\footnote{\AtlasCoordFootnote}
It consists of an inner tracking detector surrounded by a thin superconducting solenoid providing a \SI{2}{\tesla} axial magnetic field, electromagnetic and hadron calorimeters, and a muon spectrometer.
The inner tracking detector (ID) covers the pseudorapidity range \(|\eta| < 2.5\).
It consists of silicon pixel, silicon microstrip, and transition radiation tracking detectors.
Lead/liquid-argon (LAr) sampling calorimeters provide electromagnetic (EM) energy measurements with high granularity.
A steel/scintillator-tile hadron calorimeter covers the central pseudorapidity range (\(|\eta| < 1.7\)).
The endcap and forward regions are instrumented with LAr calorimeters for both the EM and hadronic energy measurements up to \(|\eta| = 4.9\).
The muon spectrometer (MS) surrounds the calorimeters and is based on three large superconducting air-core toroidal magnets with eight coils each.
The field integral of the toroids ranges between \num{2.0} and \SI{6.0}{\text{Tm}} across most of the detector.
The muon spectrometer includes a system of precision tracking chambers and fast detectors for triggering.
A two-level trigger system is used to select events.
The first-level trigger is implemented in hardware and uses a subset of the detector information to accept events at a rate below \SI{100}{\kHz}.
This is followed by a software-based trigger that reduces the accepted event rate to \SI{1}{\kHz} on average depending on the data-taking conditions.
An extensive software suite~\cite{ATL-SOFT-PUB-2021-001} is used in the reconstruction and analysis of real and simulated data, in detector operations, and in the trigger and data acquisition systems of the experiment.


\section{Data and simulated event samples}

In this search, data from proton--proton collisions at $\sqrt{s} = \SI{13}{\TeV}$ corresponding to an integrated luminosity of \SI{139}{\ifb}, collected in the years 2015 to 2018 with the ATLAS detector, are analysed.
Data are required to have been collected during stable beam conditions and with all detector subsystems operational~\cite{DAPR-2018-01}.
The average number of simultaneous $pp$ interactions per bunch crossing, referred to as \pileup, is approximately 34, averaged over the whole data set.
 
Monte Carlo (MC) simulated event samples are used to model the signal and background processes.
In all samples except those produced with \SHERPA[2.2.1] or \SHERPA[2.2.2]~\cite{Bothmann:2019yzt}, decays of heavy-flavour hadrons were modelled with \EVTGEN[1.2.0] or \EVTGEN[1.6.0]~\cite{Lange:2001uf}, depending on the process.
\Pileup was modelled by overlaying minimum-bias events generated with \PYTHIA[8.186]~\cite{Sjostrand:2007gs} and the A3~\cite{ATL-PHYS-PUB-2016-017} set of tuned parameters (referred to as the `tune') onto the simulated hard-scatter events.
A reweighting procedure was applied in order to match the \pileup profile of the recorded data.
The ATLAS simulation infrastructure~\cite{SOFT-2010-01} was used to simulate the detector and its response.
Nominal SM background samples were produced with a detailed \GEANT~\cite{Agostinelli:2002hh} detector simulation, whereas a faster calorimeter simulation~\cite{SOFT-2010-01} was applied for the signal samples and systematic variations of the backgrounds.
The same offline reconstruction methods used for data were applied to the simulated samples.
Corrections were applied to the simulated events in order to match the selection efficiencies and energy and mass scales and resolutions of reconstructed simulated particles to those measured in data control samples.

Simulated events with pair-produced scalar LQs were generated at next-to-leading order (NLO) in quantum chromodynamics (QCD) with \MGNLO[2.6.0]~\cite{Alwall:2014hca} and the \NNPDF[3.0nlo] parton distribution function (PDF)~\cite{Ball:2014uwa} set with $\alphas=0.118$.
An extension of the LQ model of Ref.~\cite{Mandal:2015lca} was used, allowing flavour off-diagonal couplings.
The model is based on previous fixed-order NLO QCD calculations~\cite{Kramer:2004df,Kramer:1997hh}.
To retain information about spin correlations, the decays of LQs as well as \topquarks were handled with \MADSPIN~\cite{Artoisenet:2012st}.
\MGNLO was interfaced with \Pythia[8.230]~\cite{Sjostrand:2014zea} to model the parton shower (PS), hadronisation, and underlying event (UE).
Here and in the following, \Pythia was used with the A14 tune~\cite{ATL-PHYS-PUB-2014-021} and the \NNPDF[2.3lo]~\cite{Ball:2012cx} set of PDFs.
The coupling strength $\lambda$ was set to 0.3, leading to a signal width of approximately \SI{0.2}{\percent}.
The model parameter $\beta \in [0,1]$ modifies the coupling of LQs to leptons, such that the coupling to charged leptons is given by $\sqrt{\beta}\lambda$ and to neutrinos by $\sqrt{1-\beta}\lambda$.   
It differs from the branching fraction \br into charged leptons
because of phase-space corrections arising mainly from the large \topquark mass. The parameter $\beta$ was set to 0.5 in the simulation.
Different values for \br were achieved by reweighting the MC events according to their decay, as described in Ref.~\cite{EXOT-2017-30}.
The LQ pair-production cross-sections were obtained from the calculation of direct top-squark pair production, as this process has the same production modes, computed at approximate next-to-next-to-leading order (\NNLO) in QCD with resummation of next-to-next-to-leading logarithmic (\NNLL) soft gluon terms~\cite{Beenakker:2016lwe,Beenakker:1997ut2,Beenakker:2010nq,Beenakker:2016gmf}.
The cross-sections do not include lepton $t$-channel contributions, which are neglected in Ref.~\cite{Mandal:2015lca} and may lead to corrections at the percent level~\cite{Borschensky:2020hot}.
For this analysis, signal samples were produced separately for electrons and muons and for both up- and down-type LQs with a mass spacing of \SI{100}{\GeV} from \SI{300}{\GeV} to \SI{800}{\GeV} and from \SI{1600}{\GeV} to \SI{2500}{\GeV}, and with a finer spacing of \SI{50}{\GeV} between \SI{800}{\GeV} and \SI{1600}{\GeV} to improve the resolution around the expected mass exclusion limit.

Simulated events with pair-produced up-type vector LQs were generated at leading order (\LO) in QCD with \MGNLO[2.8.1] and the \NNPDF[3.0nlo] PDF set with $\alphas=0.118$. Decays of the vLQs and \topquarks were handled with \MADSPIN, while the PS and hadronisation were simulated with \Pythia[8.244].
The $U_1$ vLQ model in Ref.~\cite{Baker2019} is used for the muon channel, directly seeking an explanation for the various \banomalies.
Couplings to electrons are assumed to vanish in this model because of existing tight bounds from low-energy observables, mainly in the
lepton-flavour-violating sector.
Nevertheless, an extension of the model~\cite{VLQelectrons} was used in this analysis to also probe the electron channel.
The samples were produced with a coupling strength of $g_U = 3.0$, leading to a signal width of approximately \SI{11}{\percent}.
The large value of $g_U$ is motivated by a suppression of the production cross-section for additional mediators in an ultraviolet-complete
model, which might otherwise be in tension with existing LHC limits.  
The model accommodates both left- and right-handed couplings to fermions, where in the case of only left-handed couplings the coupling strengths to charged leptons and neutrinos are equal ($\beta = 0.5$).
Although only left-handed couplings were used in producing the MC samples, the analysis uses the same reweighting of MC events
as in the scalar LQ case to also probe different values of \br.
The model allows use of either the minimal (\vLQmin) or the Yang--Mills (\vLQYM) coupling scenario, where in the latter the vLQ is a heavy gauge boson
resulting in enhanced cross-sections.
Kinematic differences between the minimal and the Yang--Mills coupling scenario were found to be negligible, except for masses below about \SI{500}{\GeV}.
Samples were produced separately for the two coupling scenarios and for both muons and electrons, with a mass spacing of \SI{100}{\GeV} from \SI{300}{\GeV} to \SI{1400}{\GeV} and from \SI{2300}{\GeV} to \SI{2500}{\GeV}, and with a finer spacing of \SI{50}{\GeV} between \SI{1400}{\GeV} and \SI{2300}{\GeV}.
No higher-order cross-section computations are available for this model.
Therefore, the cross-sections computed at leading order by \MGNLO are used in the analysis.
 
Dominant background processes in the search include \ttbar, \Wjets and single \topquark production, the last being mainly associated production of a \topquark and $W$ boson ($tW$).
In addition, \ttV ($V = W,Z$), diboson, \ttH, and \Zjets processes are also considered in the analysis.
Contributions from multi-jet background with a jet misidentified as a lepton are negligible in the phase space of interest.
 
The production of \ttbar\ events was modelled using the \POWHEGBOX~\cite{Frixione:2007nw,Nason:2004rx,Frixione:2007vw,Alioli:2010xd}~v2 generator at \NLO with the \NNPDF[3.0nlo] set of PDFs and the \hdamp parameter\footnote{The \hdamp parameter is a resummation damping factor and one of the parameters that controls the matching of \POWHEG matrix elements to the PS and thus effectively regulates the high-\pt radiation against which the \ttbar system recoils.} set to $1.5\,m_{t}$~\cite{ATL-PHYS-PUB-2016-020}.
The cross-section was corrected to the theory prediction at \NNLO including resummation of \NNLL soft-gluon terms calculated using \TOPpp[2.0]~\cite{Czakon:2011xx}.
The events were interfaced to \PYTHIA[8.230] to model the PS, hadronisation, and UE. 
 
The associated production of a \topquark and a $W$ boson was modelled using the \POWHEGBOX[v2] generator at \NLO in QCD using the five-flavour scheme and the \NNPDF[3.0nlo] set of PDFs.
The diagram removal scheme~\cite{Frixione:2008yi} was used to avoid overlap with \ttbar production because of interference.
Single-top $t$-channel ($s$-channel) production was modelled using the \POWHEGBOX[v2] generator at \NLO in QCD using the four-flavour (five-flavour) scheme and the corresponding \NNPDF[3.0nlo] PDF set.
The events were interfaced with \PYTHIA[8.230] in all cases, except for $tW$ events with large missing transverse momenta, which were interfaced with \PYTHIA[8.235].
 
The production of \Vjets was simulated with the \SHERPA[2.2.1] generator using \NLO-accurate matrix elements for up to two jets, and \LO-accurate matrix elements for up to four jets, calculated with the Comix~\cite{Gleisberg:2008fv} and \OPENLOOPS~\cite{Cascioli:2011va,Denner:2016kdg} libraries.
They were matched with the \SHERPA PS~\cite{Schumann:2007mg} using the \MEPSatNLO prescription~\cite{Hoeche:2011fd,Hoeche:2012yf,Catani:2001cc,Hoeche:2009rj} using the set of tuned parameters developed by the \SHERPA authors.
The \NNPDF[3.0nnlo] set of PDFs was used and the samples were normalised to a \NNLO prediction~\cite{Anastasiou:2003ds}.
 
Samples of diboson final states ($VV$) were simulated with the \SHERPA[2.2.1] or 2.2.2 generator, depending on the process, including off-shell effects and Higgs boson contributions, where appropriate.
Fully leptonic final states and semileptonic final states, where one boson decays leptonically and the other hadronically, were generated using matrix elements at \NLO accuracy in QCD for up to one additional parton and at \LO accuracy for up to three additional parton emissions.
The matrix element calculations were matched and merged with the \SHERPA PS.
The \NNPDF[3.0nnlo] set of PDFs was used, along with the internal \SHERPA tune.
 
The production of \ttW and \ttZ events was modelled using the \MGNLO[2.3.3] generator at \NLO with the \NNPDF[3.0nlo] PDF set.
The events were interfaced to \PYTHIA[8.210].
 
The production of \ttH events was modelled using the \POWHEGBOX[v2] generator at \NLO with the \NNPDF[3.0nlo] PDF set, interfaced to \PYTHIA[8.230].
 
An overview of the matrix element (ME) generator, PDF, shower generator and UE tune for signal and background samples is given in Table~\ref{tab:samples:generators}.

\begin{table}
\centering
\caption{List of ME generator and the order of the strong coupling constant in the perturbative calculation, PDF, shower generator and tune for the different signal and background processes.}
\label{tab:samples:generators}
\resizebox{\linewidth}{!}{\begin{tabular}{lccccc}
\toprule
Process 			& ME generator 		& ME order 		& PDF set	& PS and hadronisation 	& UE tune   \\
\midrule
Scalar LQ 			& \MGNLO[2.6.0]     	& NLO 			& \NNPDF[3.0nlo]     & \Pythia[8.230]    	& A14       \\
Vector LQ 			& \MGNLO[2.8.1]		& LO 			& \NNPDF[3.0nlo]     & \Pythia[8.244] 	& A14       \\
\ttbar/\singletop	        & \POWHEGBOX[v2] 	& NLO 			& \NNPDF[3.0nlo] 	& \PYTHIA[8.230/8.235]	& A14 	    \\
\Vjets 				& \SHERPA[2.2.1]     	& \MEPSatNLO	        & \NNPDF[3.0nnlo]	& \SHERPA 		& internal  \\
Diboson 			& \SHERPA[2.2.1/2.2.2]   & \MEPSatNLO	        & \NNPDF[3.0nnlo]    & \SHERPA		& internal  \\
\ttV 				& \MGNLO[2.3.3] 	& NLO 			& \NNPDF[3.0nlo]     & \PYTHIA[8.210]	& A14 	    \\
\ttH 				& \POWHEGBOX[v2] 	& NLO 			& \NNPDF[3.0nlo]     & \PYTHIA[8.230]    	& A14       \\
\bottomrule
\end{tabular}}
\end{table}


\section{Event reconstruction}

Events studied in this analysis are required to have at least one reconstructed $pp$ interaction vertex with at least two associated tracks with transverse momentum $\pT > \SI{0.5}{\GeV}$.
The primary vertex is selected as the one with the largest sum of squared transverse momenta of tracks associated with the interaction vertex.
In the analysis, a set of reconstructed objects is used, consisting of electrons, muons, and jets, as well as the missing transverse momentum.
When identifying charged leptons, a staggered approach is used, where so-called baseline leptons fulfil less stringent requirements than signal leptons.
Events are required to have exactly one signal lepton without any additional baseline leptons.
 
Electron candidates are reconstructed from energy deposits in the EM calorimeter matched to charged-particle tracks in the inner detector (ID).
Requirements of $|d_0|/\sigma_{d_0} < 5$ on the transverse impact parameter $d_0$ (with uncertainty $\sigma_{d_0}$) and $|\Delta z_0 \sin \theta| < \SI{0.5}{mm}$ on the longitudinal track impact parameter $z_0$ ensure matching between track and vertex.
Furthermore, electron candidates are required to lie within a pseudorapidity range of $|\eta| < 2.47$, excluding the EM calorimeter barrel--endcap transition region $1.37 < |\eta| < 1.52$.
Baseline electrons must have $\pT > \SI{10}{\GeV}$ and fulfil loose identification criteria, using a likelihood-based discriminant that combines information about tracks in the ID and energy deposits in the calorimeter system~\cite{EGAM-2018-01}.
In addition, baseline electrons are required to have a hit in the innermost layer of the pixel detector.
Isolation requirements in both the calorimeter and the ID are imposed~\cite{EGAM-2018-01}.
Electron candidates are rejected if the scalar sum of transverse momenta of tracks within a cone of size $\Delta R = \min\left(\SI{10}{\GeV}/\pt, 0.2\right)$, excluding the electron itself, is larger than \SI{15}{\percent} of the electron \pt.
Similarly, an electron is removed if, after subtracting contributions from \pileup and the electron itself, the transverse energy deposited in the calorimeter within a cone of size $\Delta R = 0.2$ exceeds \SI{20}{\percent} of the transverse energy of the electron.
To suppress backgrounds due to hadrons misidentified as electrons, signal electrons must in addition pass the `tight' identification working point, and have $\pT > \SI{30}{\GeV}$.
 
Muon candidates are reconstructed from charged-particle tracks in the ID and the MS and from energy deposits in the calorimeters.
For the reconstruction of muon candidates, tracks in the ID combined with tracks in the MS are used in the range $|\eta| < 2.5$.
In addition, 
muons in the range $|\eta| < 0.1$ are reconstructed from ID tracks matched to an energy deposit in the calorimeter
compatible with a minimally ionising particle.
These muon candidates are called calorimeter-tagged (CT).
Track-to-vertex matching is ensured by requiring $|d_0|/\sigma_{d_0} < 3$ for the transverse impact parameter and $|\Delta z_0 \sin \theta| < \SI{0.5}{mm}$ for the longitudinal track impact parameter.
Baseline muons are required to have $\pT > \SI{10}{\GeV}$ and to have compatible individual measurements in the ID and the MS.
Signal muons are required to have $\pT > \SI{30}{\GeV}$ and CT muons are not accepted~\cite{MUON-2018-03}.
Additionally, signal muons are rejected if the scalar sum of transverse momenta of tracks within a cone of size $\Delta R = \min\left(\SI{10}{\GeV}/\pt, 0.3\right)$ around the muon exceeds \SI{6}{\percent} of its transverse momentum.
In accordance with other searches for pair-produced LQs within ATLAS~\cite{EXOT-2019-13, EXOT-2019-19}, signal muons above a \pT threshold of \SI{800}{\GeV} must fulfil stricter requirements on the number of hits in the MS to ensure good momentum resolution.
 
Small-radius (small-$R$) jet candidates are built from particle-flow objects~\cite{PERF-2015-09,JETM-2018-05}, using the \antikt algorithm~\cite{Cacciari:2008gp, Fastjet} with a radius parameter of $R=0.4$.
The particle-flow algorithm combines information about ID tracks and energy deposits in the calorimeters to form the input for jet reconstruction.
Jets with $\pt < \SI{25}{\GeV}$ or $|\eta| > 2.5$ are rejected.
To reduce contributions from \pileup, jet candidates with $|\eta| < 2.4$ and $\pT < \SI{60}{\GeV}$ are required to satisfy the `tight' jet vertex tagger criterion~\cite{ATLAS-CONF-2014-018}.
Small-$R$ jets are categorised as \btagged if they satisfy a requirement on the output of a multivariate algorithm, 
operating at a tagging efficiency of \SI{77}{\percent} as determined with simulated \ttbar events~\cite{FTAG-2018-01,ATL-PHYS-PUB-2017-013}.
The $R=0.4$ jets are then reclustered with the \antikt algorithm with $R=1.0$ to obtain large-$R$ jets.
Additionally, $R=0.4$ jets are reclustered iteratively with the recursive method described in Ref.~\cite{SUSY-2016-16} to reconstruct hadronically decaying top-quark candidates ($t_{\text{had}}$).
For this, the small-$R$ jets are reclustered with an initial radius parameter of $R=3.0$, which is iteratively reduced to $R(\pt)=2m_{\text{top}}/\pt$ to match the jet radius to the top candidate's transverse momentum.
Top candidates losing large fractions of their \pt in the shrinking process are discarded.
Finally, only the leading-\pt candidate with a mass larger than \SI{150}{\GeV} is kept.
 
The missing transverse momentum (with magnitude \met) in an event is defined as the negative vectorial sum of the transverse momenta of all calibrated objects~\cite{PERF-2016-07}.
It also includes an additional track-based soft term taking into account energy depositions not associated with any calibrated object.
 
An overlap removal procedure is applied to avoid ambiguities when reconstructing the objects described above, using the baseline lepton definitions.
Electron--muon overlap is handled by removing any calorimeter-tagged muons sharing a track in the ID with an electron, and then removing any electrons sharing an ID track with a remaining muon.
Subsequently, overlap between jets and leptons is removed by rejecting any jets within $\Delta R = 0.2$ of an electron and afterwards rejecting any electrons within $\Delta R = 0.4$ of a jet.
Similarly, jets are discarded if they have fewer than three associated tracks and are within $\Delta R = 0.2$ of a muon candidate.
Otherwise, the muon is rejected if it lies within $\Delta R = \text{min}(0.4, 0.04 + \SI{10}{\GeV}/\pt(\mu))$ of a jet.


\section{Event selection and categorisation}

Any event considered in this analysis must pass an \met trigger~\cite{TRIG-2019-01}, as single-lepton triggers are found to be less efficient in particular in the muon channel.
The \met trigger thresholds varied between \SI{70}{\GeV} and \SI{110}{\GeV} across the different data-taking periods. A requirement of $\met > \SI{250}{\GeV}$ is imposed in the offline event selection to ensure full efficiency.
Events are required to contain exactly one signal lepton.
Additionally, a veto is applied on further baseline leptons.
Since the final states of interest contain one hadronically decaying \topquark and one additional jet, only events with at least four small-$R$ jets are selected.
Only one of those jets needs to be \btagged to preserve high efficiency for the signal.
To suppress contributions from fake \met caused by mismeasurements of jets or leptons, events with $\mtw~{=}~\sqrt{2\pT(\ell)\MET(1-\cos \Delta \phi(\pT(\ell), \MET))}~{<}~\SI{30}{\GeV}$ or $\Delta \phi (\met, j_{1,2}) < 0.4$, where $j_{1,2}$ indicates the highest-\pT and the second-highest-\pT small-$R$ jet, respectively, are rejected.
 
\begin{figure}
\centering
\begin{subfigure}[b]{0.5\textwidth}
\centering
\includegraphics[width=0.8\textwidth]{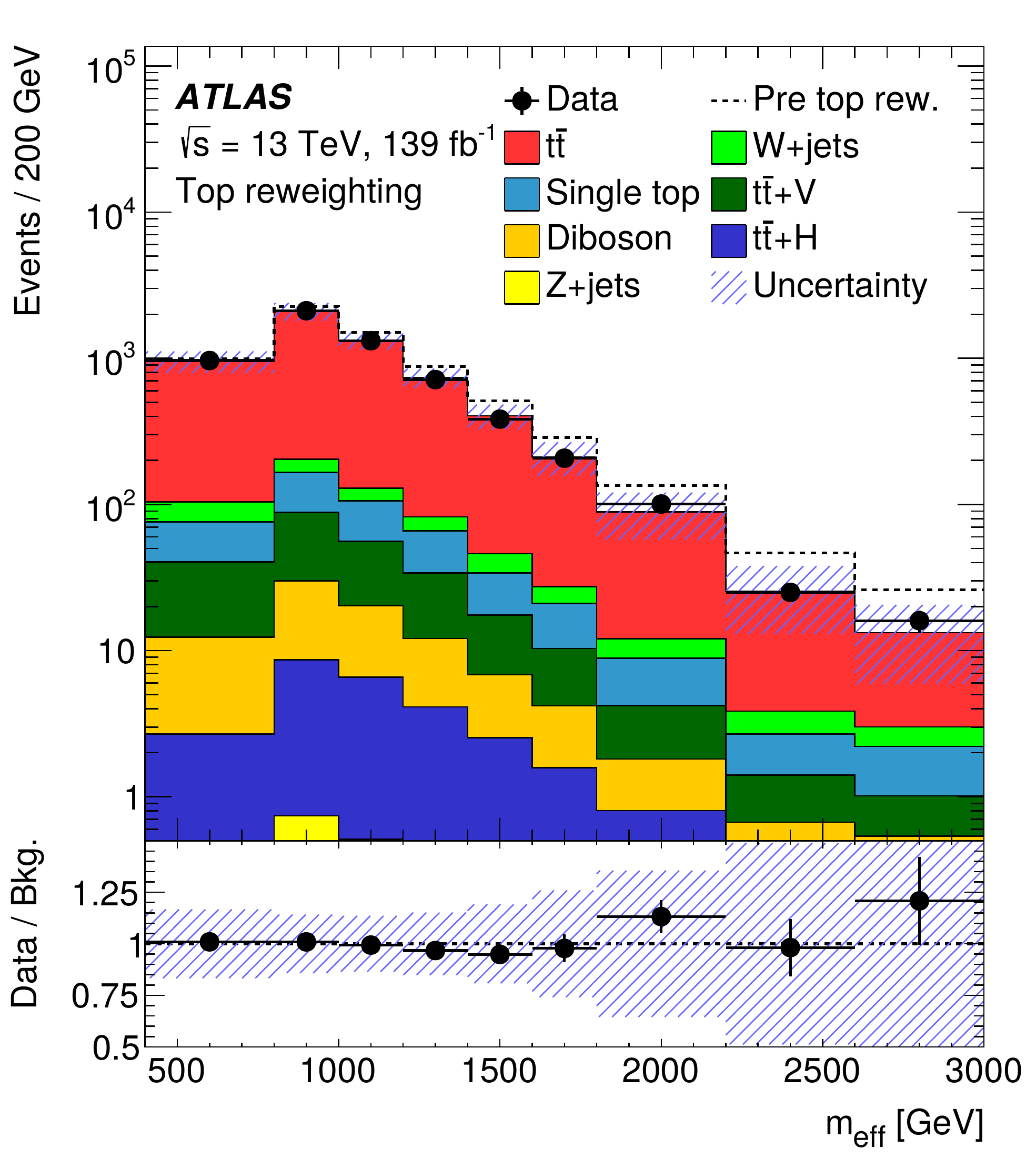}
\caption{}
\label{fig:analysis:topreweighting:meff}
\end{subfigure}\begin{subfigure}[b]{0.5\textwidth}
\centering
\includegraphics[width=0.8\textwidth]{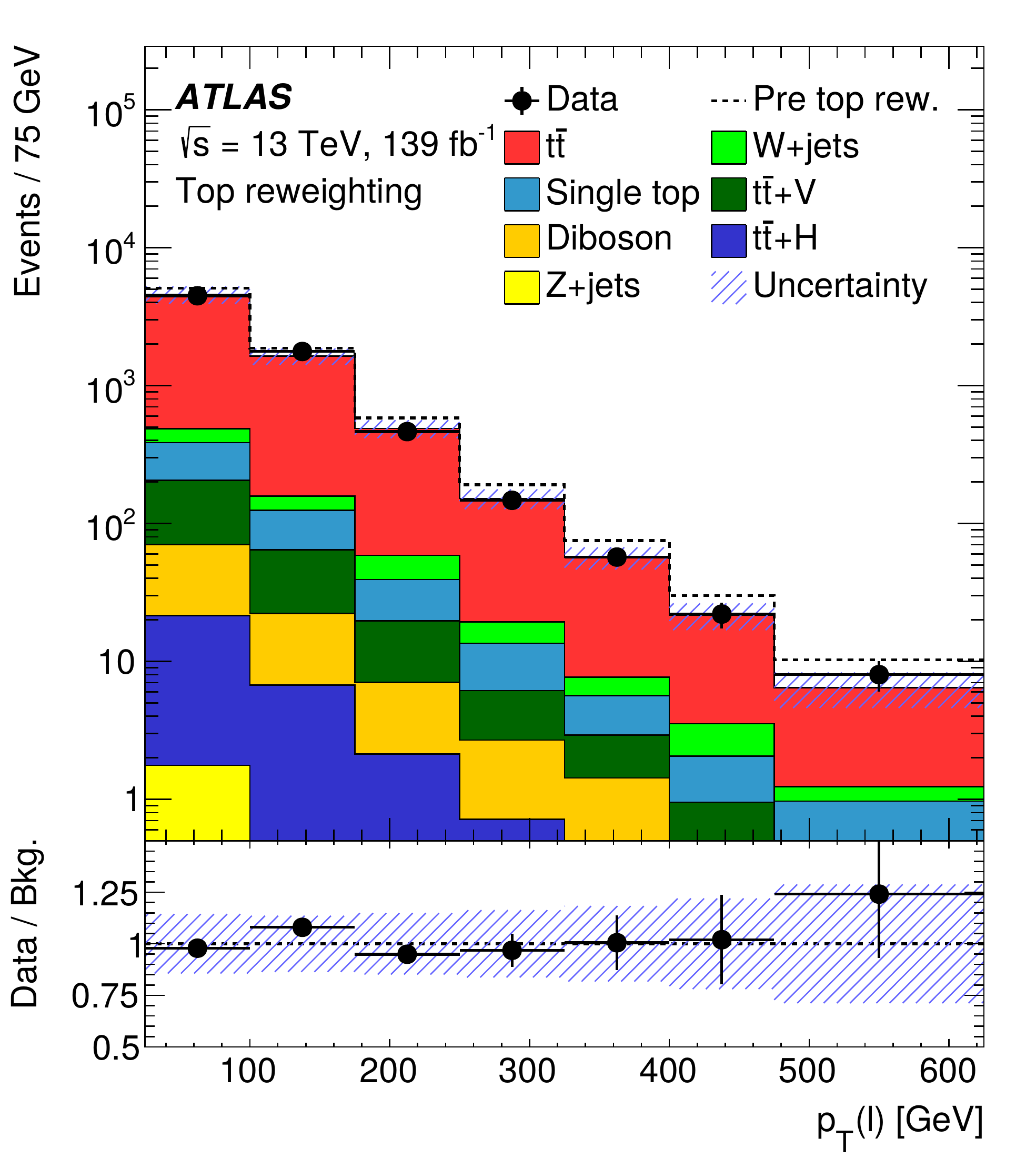}
\caption{}
\label{fig:analysis:topreweighting:leptonpt}
\end{subfigure}
\caption{Distributions of \subref{fig:analysis:topreweighting:meff} $\meff$ and \subref{fig:analysis:topreweighting:leptonpt} $\pT(\ell)$ in the \topreweighting region after applying the \topreweighting.
The hatched bands include statistical and systematic uncertainties. The total background expectation before applying the top reweighting is shown as a dashed line. The ratios of the observed and expected numbers of background events are shown in the bottom panels. The last bin contains the overflow.}
\label{fig:analysis:topreweighting}
\end{figure}

The \ttbar background is known not to be modelled accurately at high transverse momenta~\cite{TOPQ-2018-17,TOPQ-2018-15}. Reweighting factors are derived in bins of the jet multiplicity as a function of \meff, which is defined as the scalar sum of the transverse momenta of all reconstructed objects and the \met and corresponds approximately to the mass of the LQ pair.
This procedure is referred to as `top reweighting' in the following.
The reweighting factors are determined for the sum of the \ttbar and single-top backgrounds and are parameterised with a linear function, separately in each of the four jet multiplicity bins (4, 5, 6, $\geq 7$).
For this, a dedicated \topreweighting region is defined with $\amttwo < \SI{200}{\GeV}$, where the asymmetric transverse mass, \amttwo, is a variant of \mttwo~\cite{mt2Lester} and allows the reconstruction of dileptonic \ttbar events in which only one lepton is reconstructed~\cite{amt2Lester, amt2Yang}. 
The reweighting factor is then applied to single-top and \ttbar events in each of the training and control regions defined in the following.
\Cref{fig:analysis:topreweighting} shows the \meff and lepton-$\pT$ distributions in the \topreweighting region after applying the reweighting procedure, in addition to the total background expectation before reweighting.
The modelling of kinematic variables is improved by the correction.
A potential signal contribution is negligible in this region as it amounts to, e.g., \SI{0.5}{\percent} in the last four bins in \Cref{fig:analysis:topreweighting:meff} for a scalar LQ with a mass of \SI{1}{\TeV} and $\br = 0.5$.
 
\begin{figure}
\centering
\begin{subfigure}[b]{0.5\textwidth}
\centering
\includegraphics[width=0.8\textwidth]{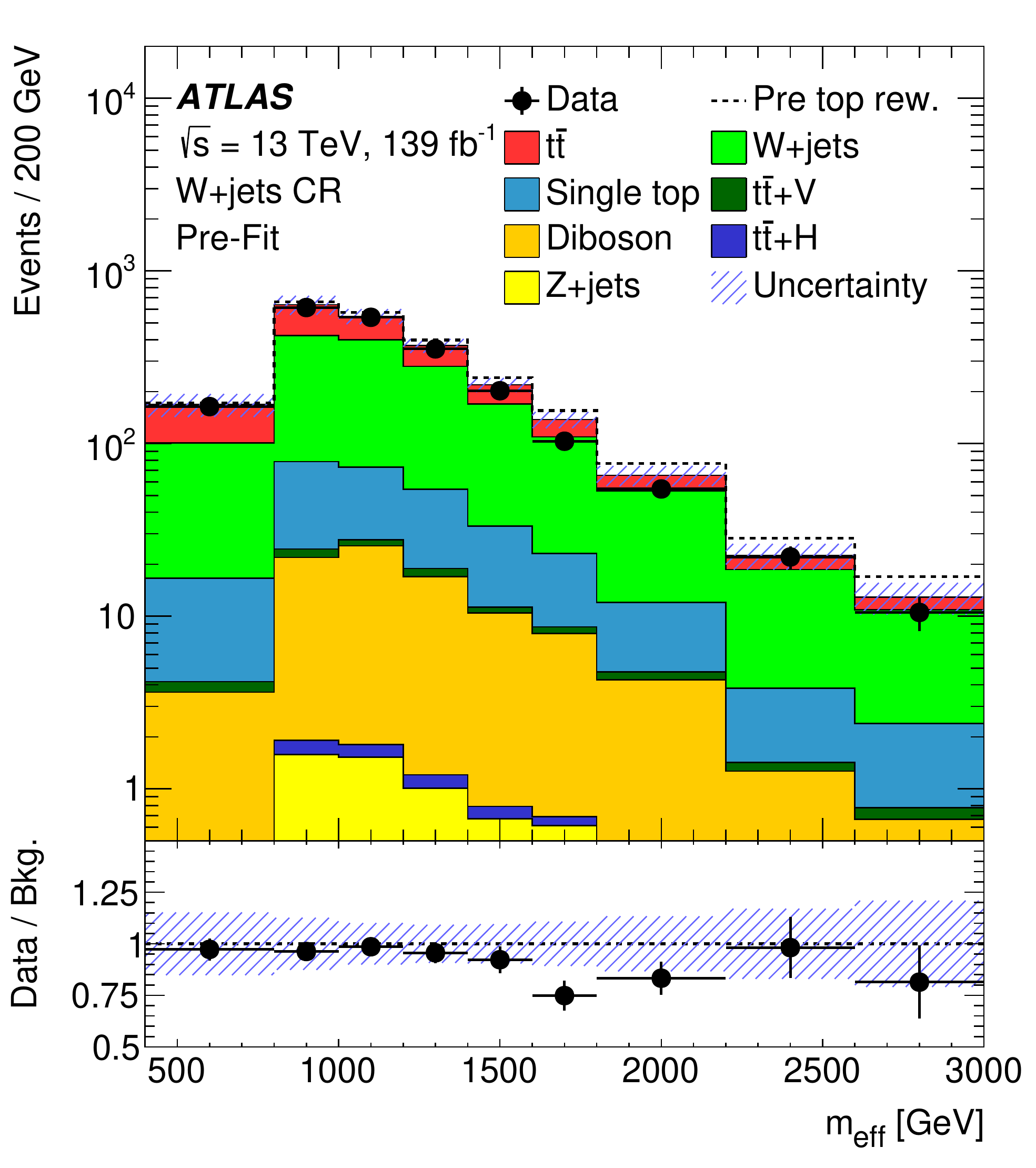}
\caption{}
\label{fig:analysis:controlregions:wjetsCR}
\end{subfigure}\begin{subfigure}[b]{0.5\textwidth}
\centering
\includegraphics[width=0.8\textwidth]{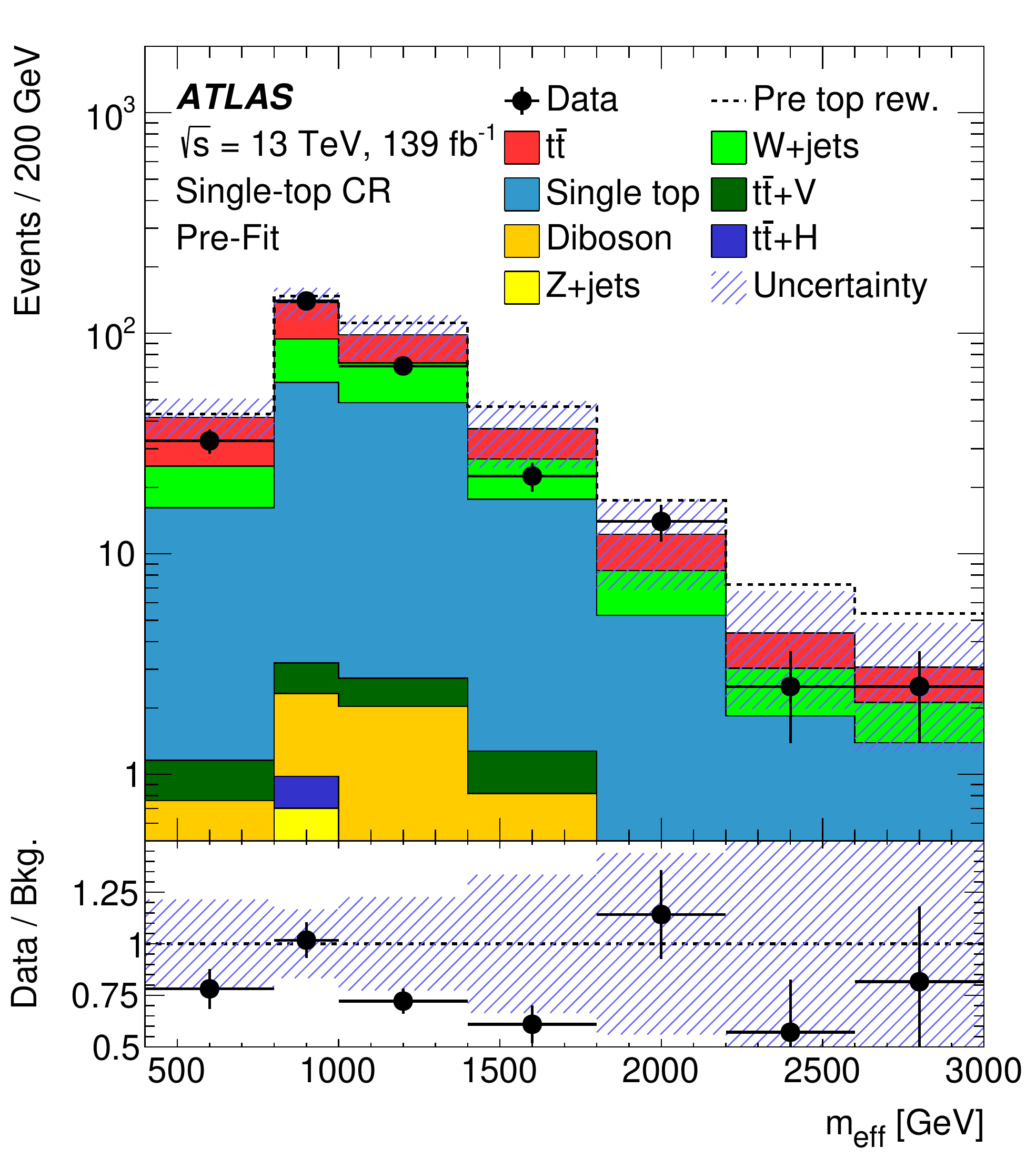}
\caption{}
\label{fig:analysis:controlregions:singletopCR}
\end{subfigure}
\caption{Distributions of $\meff$ in \subref{fig:analysis:controlregions:wjetsCR} the \Wjets CR and \subref{fig:analysis:controlregions:singletopCR} the single-top CR after applying the \topreweighting,
before the fit to data in CRs and SR.
The hatched bands include statistical and systematic uncertainties. The total background expectation before applying the top reweighting is shown as a dashed line. The ratios of the observed and expected numbers of background events are shown in the bottom panels. The last bin contains the overflow.}
\label{fig:analysis:controlregions}
\end{figure}

Control regions (CRs) enriched in the various backgrounds and with negligible signal contamination are defined. They are orthogonal to the \topreweighting region
by requiring $\amttwo > \SI{200}{\GeV}$ and are orthogonal to each other.
The \Wjets CR uses events in a window around the Jacobian peak, i.e. $\SI{50}{\GeV} \leq \mtw < \SI{120}{\GeV}$,
with a \bjet multiplicity of $n_b = 1$ and no hadronically decaying \topquark candidate.
To increase the purity of selected \Wjets events, only events with a positively charged lepton are considered, because the cross-section for $W^+$ production is larger than
that for $W^-$ production in $pp$ collisions.
For the single-top CR a requirement of $\mtw < \SI{120}{\GeV}$ is imposed.
In order to reduce contributions from \Wjets production, events must have exactly two \btagged jets, with an angular separation $\Delta R(b_1, b_2) > 1.2$.
Events containing a large-$R$ jet are vetoed.
The purity of \Wjets and single-top events in their respective CRs is \SI{58}{\percent} and \SI{38}{\percent}.
Distributions of $\meff$ in these CRs are shown in~\Cref{fig:analysis:controlregions}.
 
For the training of the NNs, a training region is defined so as to be orthogonal to the \topreweighting region and both control regions by requiring $\amttwo > \SI{200}{\GeV}$ and $\mtw \geq \SI{120}{\GeV}$.
An overview of the different selection requirements for the training and control regions is given in Table~\ref{tab:analysis:selection}.
The product of signal acceptance and efficiency in the training region is similar for signal hypotheses with couplings to electrons or muons. For $\br = 0.5$, it amounts to around \SI{17}{\percent} for up-type scalar LQs and to around \SI{14}{\percent} for down-type scalar LQs at $\mlq = \SI{1.4}{\TeV}$.
For vector LQs, it reaches \SI{20}{\percent} at the same mass for both the Yang--Mills coupling and the minimal coupling scenario.
 
\begin{table}
\caption{Overview of event selections applied in the different regions of the analysis.}
\label{tab:analysis:selection}
\centering
\resizebox{\textwidth}{!}{\begin{tabular}{cccc}
\toprule
\multicolumn{4}{c}{Preselection} \\
\midrule
\multicolumn{4}{c}{\met triggers} \\
\multicolumn{4}{c}{exactly one signal lepton} \\
\multicolumn{4}{c}{veto on additional baseline leptons} \\
\multicolumn{4}{c}{$\met > \SI{250}{\GeV}$} \\
\multicolumn{4}{c}{$\geq 4$ small-$R$ jets} \\
\multicolumn{4}{c}{$\mtw > \SI{30}{\GeV}$} \\
\multicolumn{4}{c}{$\Delta \phi (\met, j_{1,2}) > 0.4$} \\
\midrule
Top reweighting region  & \Wjets CR & Single-top CR & Training region   \\
&           &               & Low-\nnout CR/SR  \\
\midrule
$n_b \geq 1$ & $n_b = 1$ & $n_b = 2$ & $n_b \geq 1$ \\
$\mtw \geq \SI{120}{\GeV}$ & $\SI{50}{\GeV} \leq \mtw < \SI{120}{\GeV}$ & $\mtw < \SI{120}{\GeV}$ & $\mtw \geq \SI{120}{\GeV}$ \\
$\amttwo < \SI{200}{\GeV}$ & $\amttwo > \SI{200}{\GeV}$ & $\amttwo > \SI{200}{\GeV}$ & $\amttwo > \SI{200}{\GeV}$ \\
- & $t_{\text{had}}$ candidate veto & large-$R$ jet veto & - \\
- & lepton charge $= +1e$ & - & - \\
- & - & $\Delta R(b_1, b_2) > 1.2$ & - \\
- & - & - & $\nnout < 0.5 / \geq 0.5$ \\
\bottomrule
\end{tabular}}
\end{table}


\section{Neural network training}

Simulated signal and background events in the training region are used to train several NNs for the various signal hypotheses.
The NNs are implemented using the NeuroBayes package~\cite{feindt-2004,Feindt:2006pm} which combines a
three-layer feed-forward NN with a complex and robust preprocessing of the input variables prior to their presentation to the NN.
The purpose of the preprocessing is to facilitate optimal network training by ordering the input variables according to their ability to discriminate between signal and background, taking correlations into account, and removing all but the most powerful ones.
 
NeuroBayes uses Bayesian regularisation techniques for the training process to improve the generalisation performance and to avoid overtraining.
In general, the network infrastructure consists of one input node for each input variable plus one bias node, an arbitrary, user-defined number of hidden nodes, and one output node which gives a continuous NN output score (\nnout) in the interval $(0,+1)$, where large values indicate signal-like
events and small values background-like events.
For the NNs of this analysis, 15 nodes are used in the hidden layer and the ratio of signal to background events in the training is 1:1.
The different background processes are weighted according to their expected number of events.
Only \ttbar, \Wjets, single-\topquark, and \ttV events are used as background processes in the training.
As a check for potential overtraining, only \SI{80}{\percent} of the simulated events serve as input to the training, while the remaining \SI{20}{\percent} are used as a test sample.
No signs of overtraining are observed.
After the training step, samples of simulated signal and background events, as well as the observed events, are processed by the NNs in order to
get an \nnout value for each event.
For each NN, the training region is divided into a \lownnout control region with \nnout < 0.5, enriched mainly in \ttbar events, and the signal region above 0.5.
 
\begin{figure}
\centering
\begin{subfigure}[b]{0.5\textwidth}
\centering
\includegraphics[width=0.8\textwidth]{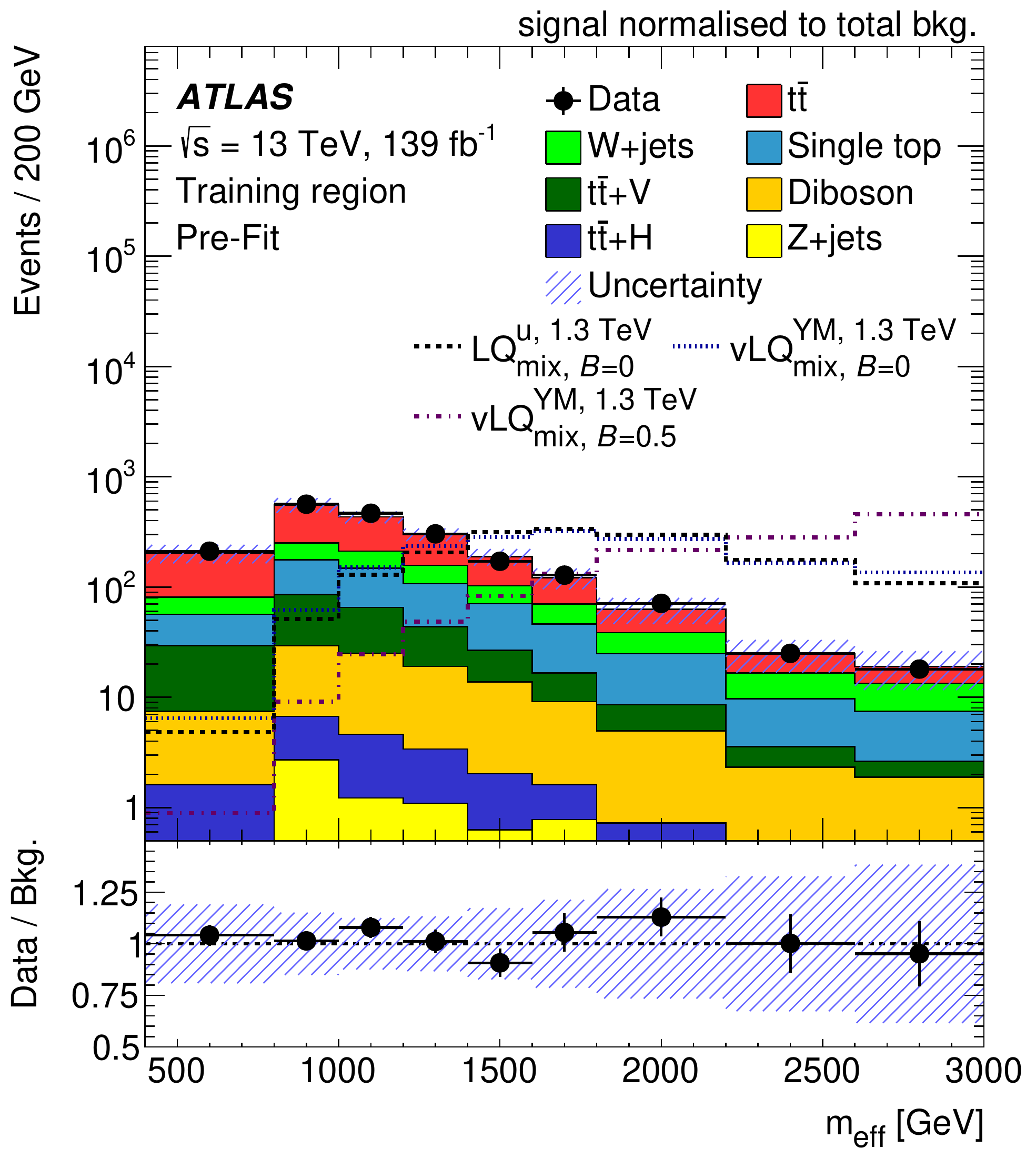}
\caption{}
\label{fig:analysis:signalregion:meff}
\end{subfigure}\begin{subfigure}[b]{0.5\textwidth}
\centering
\includegraphics[width=0.8\textwidth]{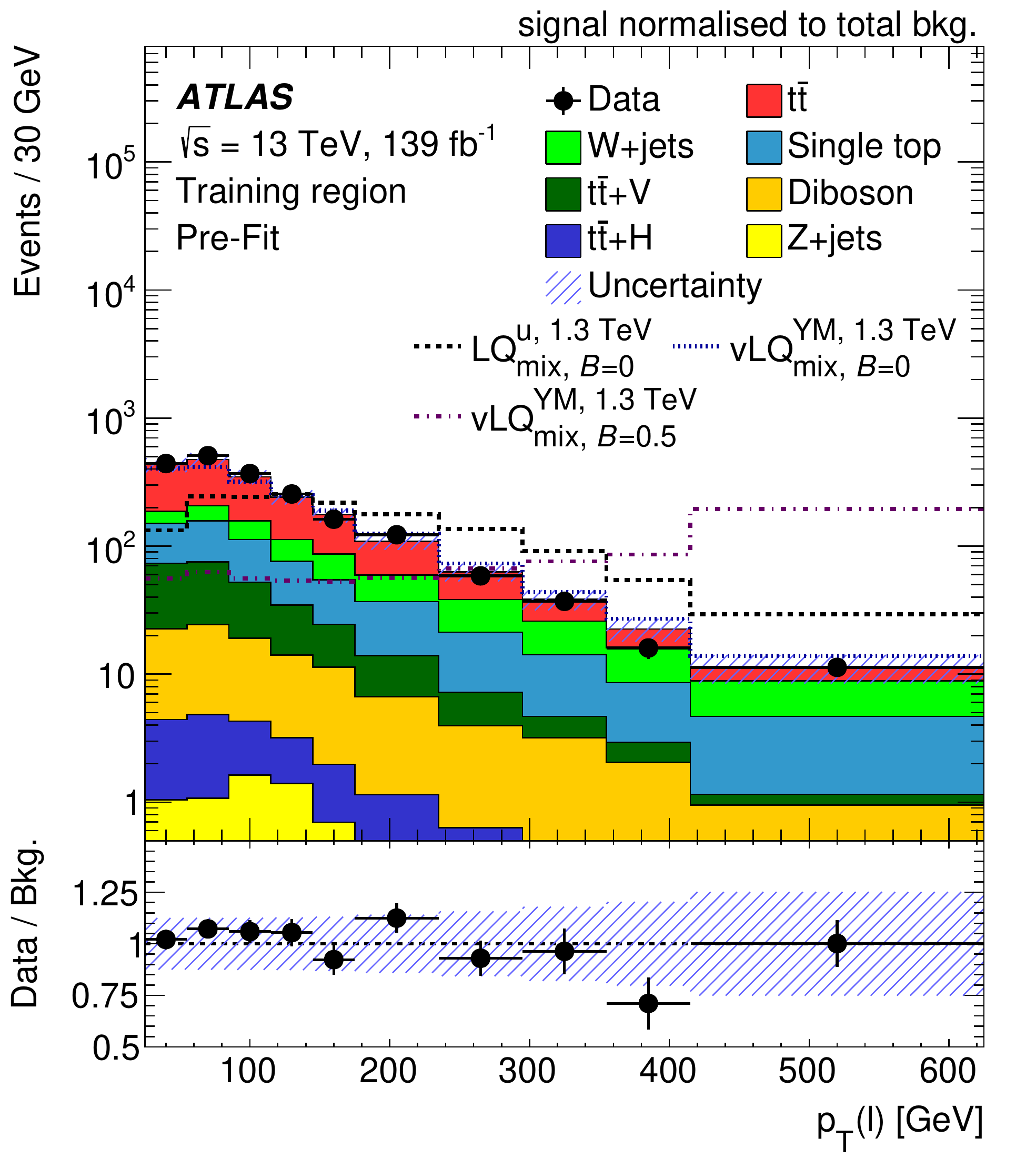}
\caption{}
\label{fig:analysis:signalregion:leptonpt}
\end{subfigure}
\begin{subfigure}[b]{0.5\textwidth}
\centering
\includegraphics[width=0.8\textwidth]{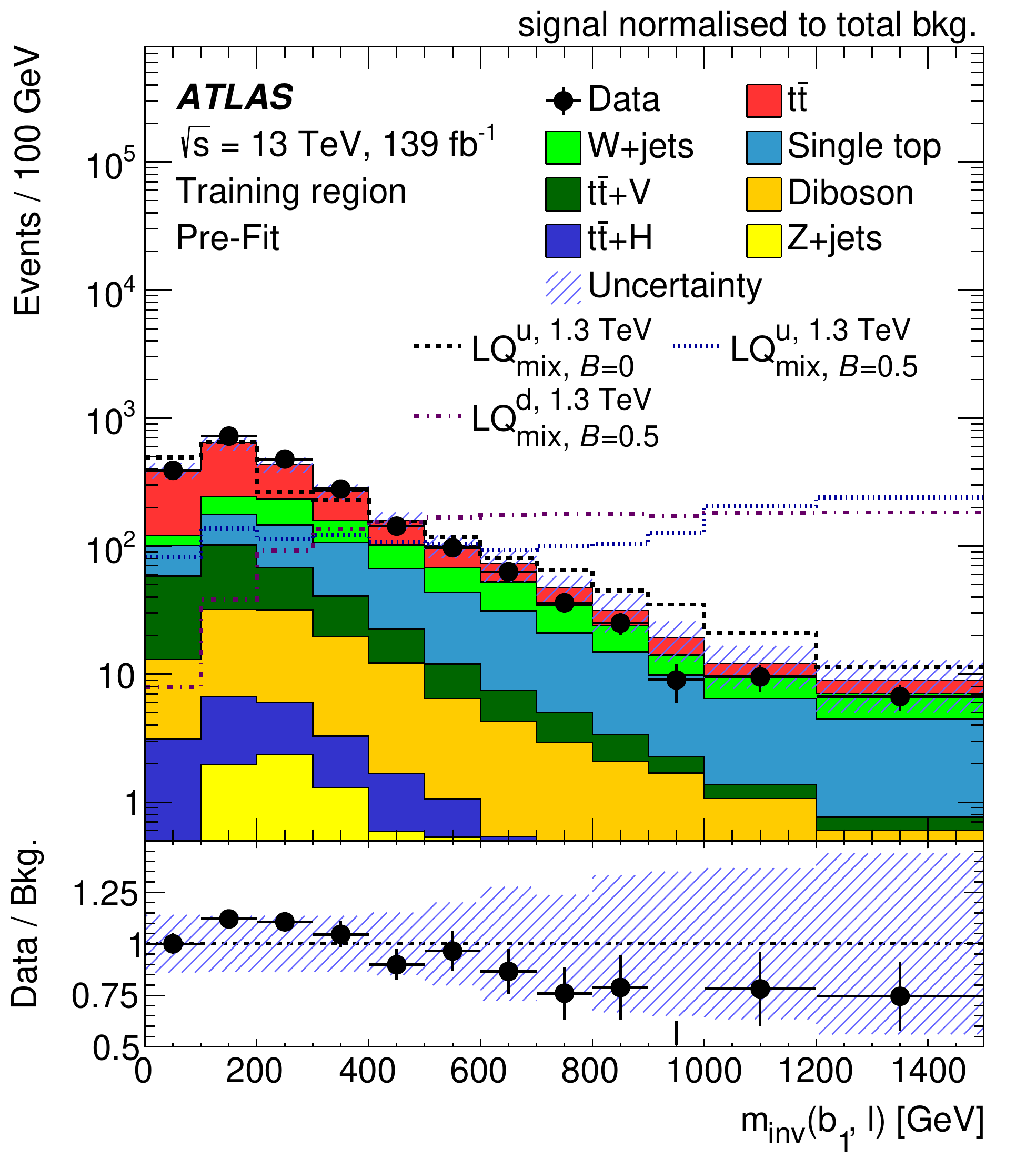}
\caption{}
\label{fig:analysis:signalregion:minvbl}
\end{subfigure}\begin{subfigure}[b]{0.5\textwidth}
\centering
\includegraphics[width=0.8\textwidth]{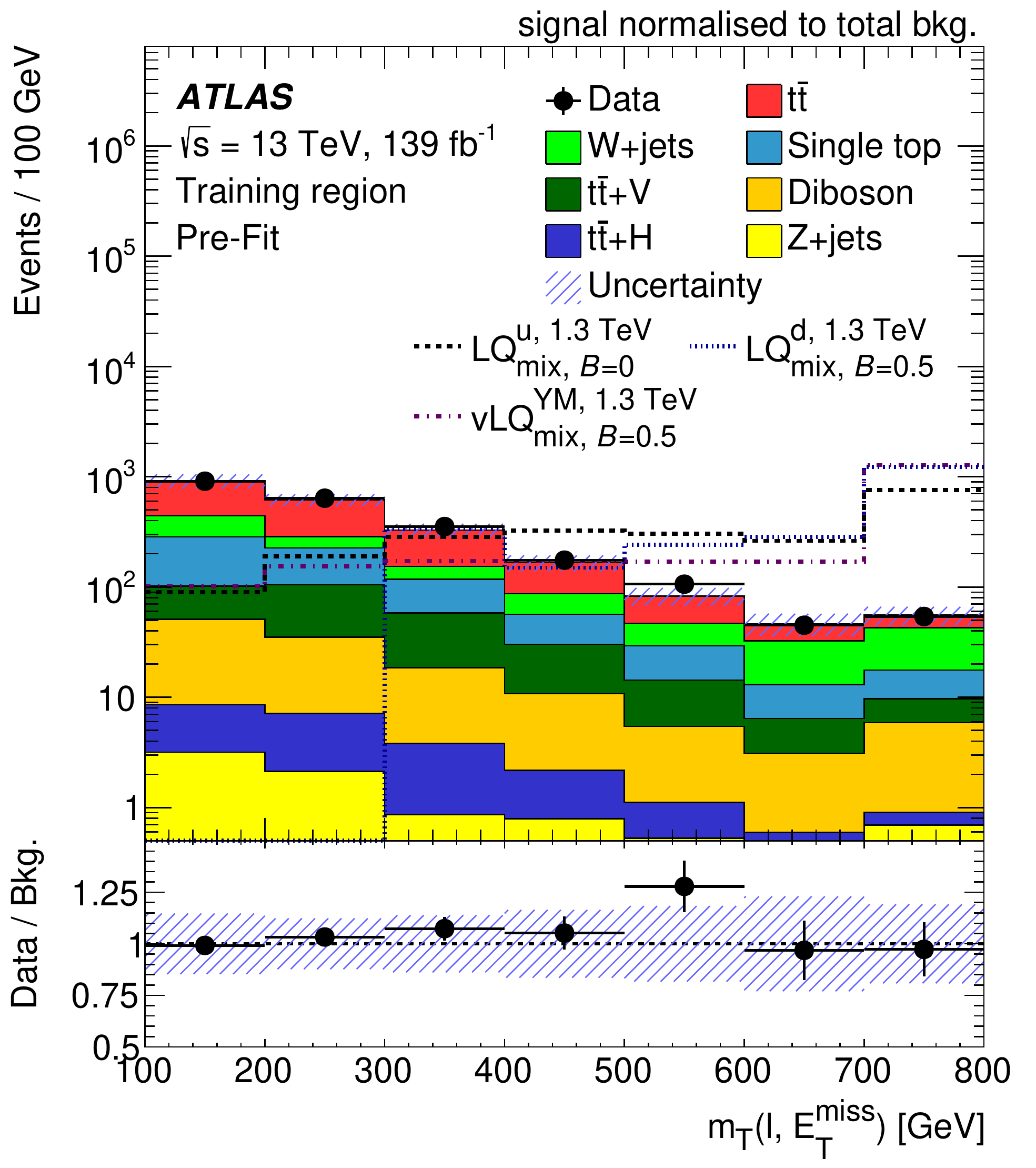}
\caption{}
\label{fig:analysis:signalregion:mtw}
\end{subfigure}
\caption{Distributions of \subref{fig:analysis:signalregion:meff} $\meff$, \subref{fig:analysis:signalregion:leptonpt} $\pT(\ell)$, \subref{fig:analysis:signalregion:minvbl} $\minv(b_1, \ell)$, and \subref{fig:analysis:signalregion:mtw} $\mtw$ in the training region after applying the \topreweighting, before the fit to data in CRs and SR. The hatched bands include statistical and systematic uncertainties.
Signal distributions normalised to the total background expectation are overlaid for
up-type scalar LQs with $\br=0.0$ and vector LQs with $\br=0.0$ and 0.5, each with $\mlq=\SI{1.3}{\TeV}$.
The ratios of the observed and expected numbers of background events are shown in the bottom panels. The last bin contains the overflow.}
\label{fig:analysis:signalregion}
\end{figure}

The input variables are chosen because of their ability to discriminate between signal and background.
In total, 15 input variables are provided for the training, including the lepton flavour in order to distinguish between electrons and muons. This is because a final state with one lepton flavour has some sensitivity to a LQ model with the other flavour if the lepton stems from a \topquark decay, i.e.\ mainly in the low \br region.
\Cref{tab:analysis:inputvariables} lists the input variables in order of decreasing ability to discriminate between signal and background. The order is not absolute, as there is some dependence on the signal
model and \br, e.g.\ the lepton flavour cannot discriminate at all in the region of low \br, but is important otherwise.
The  modelling of the input variables in the training region is good in general, as can be seen for the most important ones
in~\Cref{fig:analysis:signalregion}.            
Among the least well modelled of all input variables is $\minv(b_1, \ell)$ shown in~\Cref{fig:analysis:signalregion:minvbl}. This
is due to the interference between single-top and \ttbar production, which is difficult to describe in MC simulations~\cite{ATL-PHYS-PUB-2021-042}.
 
\begin{table}
\centering
\caption{Input variables for the NN training, approximately sorted in descending ability to discriminate
between signal and background. The order is not absolute as there is some dependence on the signal model and \br. Some variables might not be defined in every event.}
\label{tab:analysis:inputvariables}
\begin{tabular}{ll}
\toprule
Variable                            & Description \\
\midrule
$\mtw$                              & transverse mass of lepton and \met \\
$\meff$                             & scalar sum of the transverse momenta of leptons, jets, and \MET \\
Lepton flavour                      & flavour of the signal lepton \\
$\pt(\ell)$                         & transverse momentum of the lepton \\
$\minv(b_1,\ell)$                   & invariant mass of the leading-\pt \bjet and the lepton  \\
$n_\text{large}$                            & reclustered large-$R$ jet multiplicity  \\
$\amttwo$                           & asymmetric transverse mass  \\
\MET significance                   & measure for assessing the compatibility of the observed \MET with zero, \\
& taking resolutions into account  \\
$m_\text{T}(b_1, \met)$                    & transverse mass of leading-\pt \bjet and \met \\
$\pt(t_{\text{had}})$               & transverse momentum of $t_{\text{had}}$  \\
$\Delta \phi(\met, b_2)$            & azimuthal angle separation between \met and subleading-\pt \bjet \\
$\minv(b_2,\ell)$                   & invariant mass of subleading-\pt \bjet and lepton \\
$\Delta \phi(\met, b_1)$            & azimuthal angle separation between \met and leading-\pt \bjet  \\
$\Delta \phi(t_{\text{had}}, \ell)$ & azimuthal angle separation between $t_{\text{had}}$ and lepton \\
$\pt(b_1)$                          & transverse momentum of leading-\pt \bjet  \\
\bottomrule
\end{tabular}
\end{table}
 
\Cref{fig:analysis:signalregion} also displays three up-type signal hypotheses at a mass of \SI{1.3}{\TeV}. It can be seen that the signal shape depends not only on the value of \br but also on the spin of the LQ. The signal shape differences due to spin correlations are sizeable for low values of \br at
small values of lepton \pt, where the lepton usually originates from a \topquark decay. Here, the lepton-\pt distribution in the vector LQ model is found to be similar in shape to the background in contrast to the scalar case as can be seen~\Cref{fig:analysis:signalregion:leptonpt}.
Therefore, separate NNs are trained for scalar and vector LQs as well as for various values of \br.
A total of four NNs per lepton flavour at $\br=0.0, 0.25, 0.5$, and $0.9$ are used for up-type scalar and vector LQs. Since kinematic differences between the Yang--Mills coupling and the minimal coupling are negligible except for very low masses, the NNs for vector LQs are only trained on samples with the former coupling and then applied to the latter as well.
In the case of down-type scalar LQs, only one NN is trained for LQs decaying into muons and another is trained for LQs decaying into electrons. For both trainings, a branching ratio of 0.5 is assumed, since only events with one LQ decaying into a charged lepton and the other decaying into a neutrino contribute significantly to the phase space under consideration.
For scalar LQs, signal samples produced with $\mlq = \SI{500}{\GeV}, \SI{900}{\GeV}$, and \SI{1300}{\GeV} are combined in each training, as separate trainings for each signal mass did not yield a significant improvement for masses above about \SI{600}{\GeV}. For vector LQs, additional signal samples with $\mlq = \SI{1700}{\GeV}$ are used to take advantage of their higher production cross-sections and therefore higher expected mass limits when compared to scalar LQs.


\section{Systematic uncertainties}

 
The largest systematic uncertainties considered in the analysis are related to the modelling of the
background processes.
For \ttbar production, the uncertainty due to the method used to match the matrix element to the parton shower is assessed by comparing a sample produced by \MGNLO with the nominal sample from \POWHEGBOX, using the same parton shower.
Conversely, an estimate of the uncertainties related to the underlying event, parton shower, and hadronisation is obtained by
comparing a sample showered by \Herwig[7]~\cite{Bellm:2017jjp} with the nominal sample showered by \Pythia, using the same ME generator for both.
The effects of uncertainties in the renormalisation and factorisation scales are estimated by independently varying the scales by a factor of two.
The impact of initial-state radiation (ISR) is estimated by varying $\alphas$ in the A14 tune.   
Similarly, the uncertainty related to final-state radiation (FSR) is assessed by varying the renormalisation scale for final-state parton-shower emissions by a factor of two.
Additionally, an uncertainty related to the choice of value for the \POWHEGBOX-specific \hdamp parameter is evaluated by using a varied value of $\hdamp = 3.0 \, m_t$.
PDF uncertainties are obtained from the PDF4LHC15 PDF set~\cite{Butterworth:2015oua}.
 
For single top-quark production, uncertainties in ME-to-PS matching, the choice of parton shower, renormalisation and factorisation scales, ISR, FSR, and PDF are evaluated using the same procedures as for \ttbar production.
Large uncertainties arise due to interference effects between \ttbar and $tW$ production.
They are estimated by comparing the nominal sample based on the diagram removal scheme with a sample using the diagram subtraction scheme~\cite{Frixione:2008yi,Re:2010bp}.
 
In the \topreweighting procedure, the parameters of the linear fit are varied within their $2\sigma$ uncertainty to account for potential non-linearities in addition to the statistical uncertainty, treating the normalisation and the shape component in each of the four jet multiplicity bins independently.
Compared to the $1\sigma$ variation, this choice has a negligible impact on the results.
 
For \Wjets, \Zjets, diboson, \ttV, and \ttH processes, renormalisation and factorisation scale variations are considered, following the same procedures as for \ttbar.
In addition, an uncertainty of \SI{50}{\percent} is assigned to the heavy-flavour component of the \Wjets background to cover differences in flavour composition between control and signal regions seen in MC studies.
 
Theoretical systematic uncertainties also include cross-section uncertainties for those background processes for which the normalisation is not determined in the fit.
For diboson and \Zjets production, this uncertainty is taken to be \SI{6}{\percent}~\cite{ATL-PHYS-PUB-2016-002} and \SI{5}{\percent}~\cite{STDM-2015-03}, respectively.
For \ttH production, it amounts to \SI{11}{\percent}~\cite{deFlorian:2016spz} and for \ttZ production to \SI{15}{\percent}~\cite{deFlorian:2016spz}.
For \ttW production, the cross-section uncertainty is taken to be \SI{50}{\percent} to account for potential differences between predicted and measured values as reported in Ref.~\cite{TOPQ-2016-11}.
 
Systematic uncertainties in the signal prediction arise from acceptance effects due to renormalisation and factorisation scale, ISR/FSR, and PDF and $\alphas$ variations. They were found to not exceed \SI{5}{\percent} in total across the whole mass range, and this is therefore taken as a conservative estimate.
 
Additionally, detector-related uncertainties are considered, the dominant ones being the small-$R$ jet energy scale and resolution uncertainties~\cite{JETM-2018-05}.
Furthermore, systematic uncertainties related to the jet mass scale and resolution, the lepton identification, isolation, and reconstruction efficiencies as well as the lepton energy scale and resolution~\cite{EGAM-2018-01,MUON-2018-03}, the $b$-tagging efficiencies~\cite{FTAG-2018-01}, and the \met reconstruction~\cite{ATLAS-CONF-2018-023} are taken into account.
Minor contributions to the total systematic uncertainty also come from
the uncertainty of \SI{1.7}{\percent} in the integrated luminosity~\cite{ATLAS-CONF-2019-021}, obtained using the LUCID-2 detector~\cite{LUCID2} for the primary luminosity measurements, and from an uncertainty related to \pileup reweighting.
 
All sources of systematic uncertainty affect the total event yield, and all, except the ones where the size of the uncertainty is explicitly stated above, also affect the shape of the distributions used in the fit.


\section{Statistical interpretation}

The binned distributions of the NN output are used to test for the presence of a signal.
Simultaneous binned profile-likelihood fits are performed for hypothesis testing, following a modified frequentist method implemented in RooStats~\cite{RooFit} and using the \nnout distribution in the signal region and the overall number of events in the \lownnout, \Wjets, and single-top control regions.
Systematic uncertainties affecting signal and background expectations are accounted for by including them in the fit in the form of nuisance parameters.
For uncertainties in the modelling of background processes for which the normalisation is determined in the likelihood fit, only shape effects and acceptance differences between the CRs and the SR are considered, in order to avoid double-counting of normalisation uncertainties.
This procedure is also used for pre-fit uncertainty bands in order to have an equivalent treatment of systematic uncertainties.
A smoothing algorithm is applied to certain systematic variations in the signal region in order to reduce statistical fluctuations between bins.
To simplify the fitting procedure, for each region and each process a nuisance parameter is only considered if the overall effect on the normalisation of the process is larger than \SI{1}{\percent}.
In the case of the signal region, which has multiple bins, nuisance parameters are also considered if their effect in any bin within the signal region is above \SI{1}{\percent}.
 
The binned likelihood function $\mathcal{L}(\mu,\theta)$ is constructed as the product of Poisson probability terms over all bins considered in the analysis.
It depends on the signal strength parameter $\mu$, a multiplicative factor applied to the theoretical signal production cross-section, and $\theta$, a set of nuisance parameters, implemented in the likelihood function as Gaussian priors for shape effects and as log-normal priors for normalisation effects.
The expected number of events in a bin depends on $\mu$ and $\theta$.
The nuisance parameters $\theta$ adjust the expectations for signal and background according to the corresponding systematic uncertainties, and their fitted values correspond to the amounts that best fit the data.
 
The test statistic $q_{\mu}$ is defined as the profile likelihood ratio
\begin{align*}
q_{\mu} = -2 \ln{\frac{\mathcal{L}(\mu,\hat{\hat{\theta}})}{\mathcal{L}(\hat{\mu},\hat{\theta})}},
\end{align*}
where $\hat{\mu}$ and $\hat{\theta}$ are the values of the parameters that maximise the likelihood function (with the constraints $0 \le \hat{\mu} \le \mu$), and $\hat{\hat{\theta}}$ are the values of the nuisance parameters (NPs) that maximise the likelihood function for a given value of $\mu$.
This test statistic is used to determine whether the observed data are compatible with the background-only hypothesis, i.e.\ with $\mu = 0$.
Furthermore, by using the CL$_\text{s}$ method~\cite{Read:2002hq}, upper limits on the signal production cross-section are derived for each of the signal scenarios considered in this analysis.
For a given signal scenario, values of the production cross-section (parameterised by $\mu$) yielding CL$_\text{s} < 0.05$, where CL$_\text{s}$ is computed using the asymptotic approximation~\cite{Cowan:2010js}, are excluded at $\geq \SI{95}{\percent}$ confidence level (CL).


\section{Results}

 
For each NN training, a separate fit to the \nnout distribution in the signal region and the overall number of events in the \lownnout, \Wjets, and single-top control regions is performed, with free normalisation parameters for the \ttbar, single-top, and \Wjets background processes.
The normalisation parameters obtained from fits to data using the background-only hypothesis are consistent across all trainings.
They are always applied in the following and
vary between $1.09 \pm 0.22$ and $1.29 \pm 0.23$ for \ttbar, between $0.84 \pm 0.12$ and $0.93 \pm 0.12$ for \Wjets, and between $0.46 \pm 0.27$ and $0.54 \pm 0.26$ for \singletop.
The normalisation parameter for the single-top process reduces the event yield by approximately a factor of two; however, the expected yield from the alternative scheme to model the interference between the \ttbar and $tW$ processes leads to an even smaller yield.
Observed and expected event yields after the background-only fit
are listed in~\Cref{tab:results:yields} for one NN training.
 
\begin{table}
\centering
\caption{Observed and expected event yields in the control and signal regions for a training for $\vLQYM \rightarrow b\mu/t\nu$ and $\br=0.5$ after the background-only fit. The uncertainties in the background predictions include both the statistical and systematic components. For comparison, expected event yields are shown for a \vLQYM\ signal at a mass point of \SI{1700}{\GeV} and $\br = 0.5$ including its pre-fit uncertainties.}
\label{tab:results:yields}
\begin{tabular}{lcccc}
\toprule
&	\Wjets CR 		& Single-top CR 	& Low-$NN_\text{out}$ CR 	& SR \\
\midrule
\ttbar 			& $860\pm140$		& $186\pm35$~~	& $1370\pm150$~~ 			& $53\pm10$  \\
Single top 		& $103\pm87$~~		& $131\pm47$~~	& $200\pm110$				& $36\pm14$  \\
\Wjets 			& $1240\pm130$~~		& $101\pm28$~~	& $265\pm55$~~			& $32.4\pm6.9$~~  \\
\ttV 				& $11.0\pm1.8$	~    		& $4.47\pm0.79$	& $180\pm28$~~			& $16.7\pm2.6$~~  \\
Diboson			& $94.3\pm9.8$~~		& $7.6\pm1.9$		& $94\pm11$				& $11.1\pm1.2$~~  \\
\ttH 				& $1.27\pm0.16$		& $1.00\pm0.12$	& $14.4\pm1.7$~~			& $1.34\pm0.18$  \\
\Zjets 			& $6.46\pm0.32$ 		& $2.18\pm0.11$	& $7.20\pm0.36$			& $1.31\pm0.07$  \\
\midrule
Total background 	& $2308 \pm 48$~~~~ 	& $433 \pm 21$~~	& $2126 \pm 46$~~~~		& $152 \pm 13$~~  \\
Observed events  	& 2310 				& 430 			& 2124 					& 157  \\
\midrule
\vLQYM\,(\SI{1.7}{\TeV}, $\br=0.5$)	& $0.109 \pm 0.022$ & $0.097 \pm 0.016$ & $1.57 \pm 0.10$	& $38.9 \pm 2.6$~~ \\
\bottomrule
\end{tabular}
\end{table}

\begin{figure}
\centering
\begin{subfigure}[b]{0.33\textwidth}
\centering
\includegraphics[width=\textwidth]{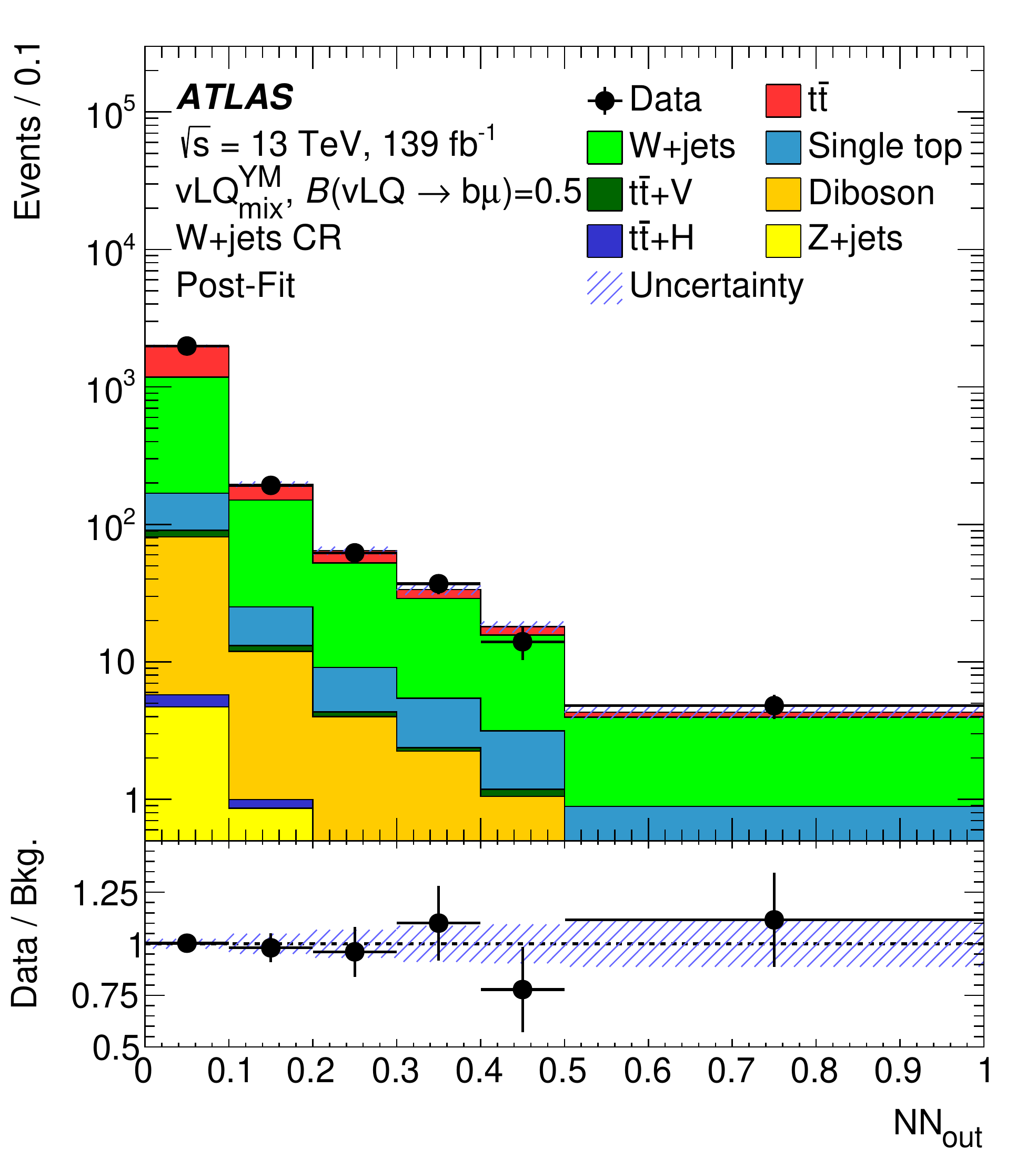}
\caption{}
\label{fig:results:validation:wjetsCR}
\end{subfigure}\begin{subfigure}[b]{0.33\textwidth}
\centering
\includegraphics[width=\textwidth]{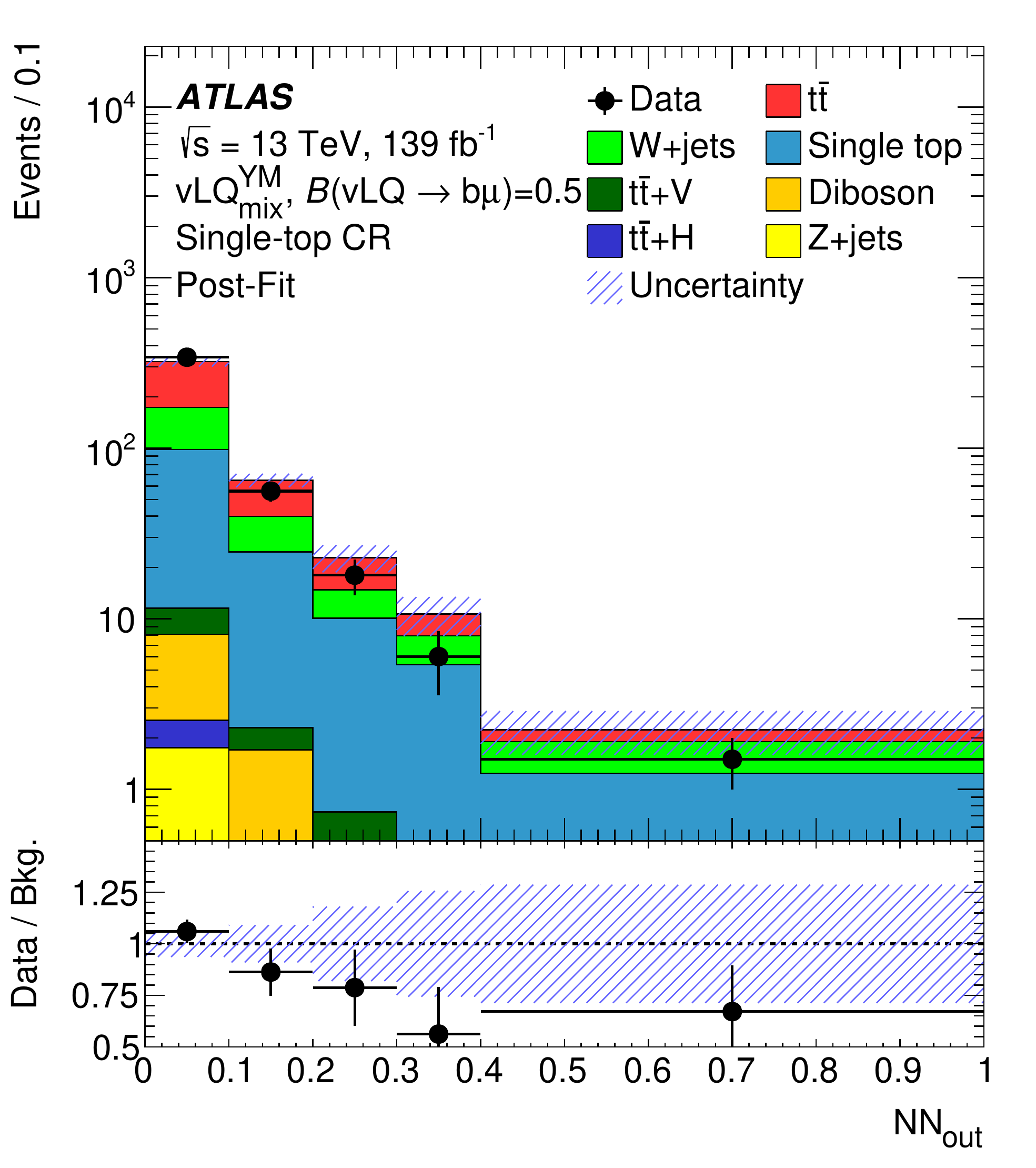}
\caption{}
\label{fig:results:validation:singletopCR}
\end{subfigure}\begin{subfigure}[b]{0.33\textwidth}
\centering
\includegraphics[width=\textwidth]{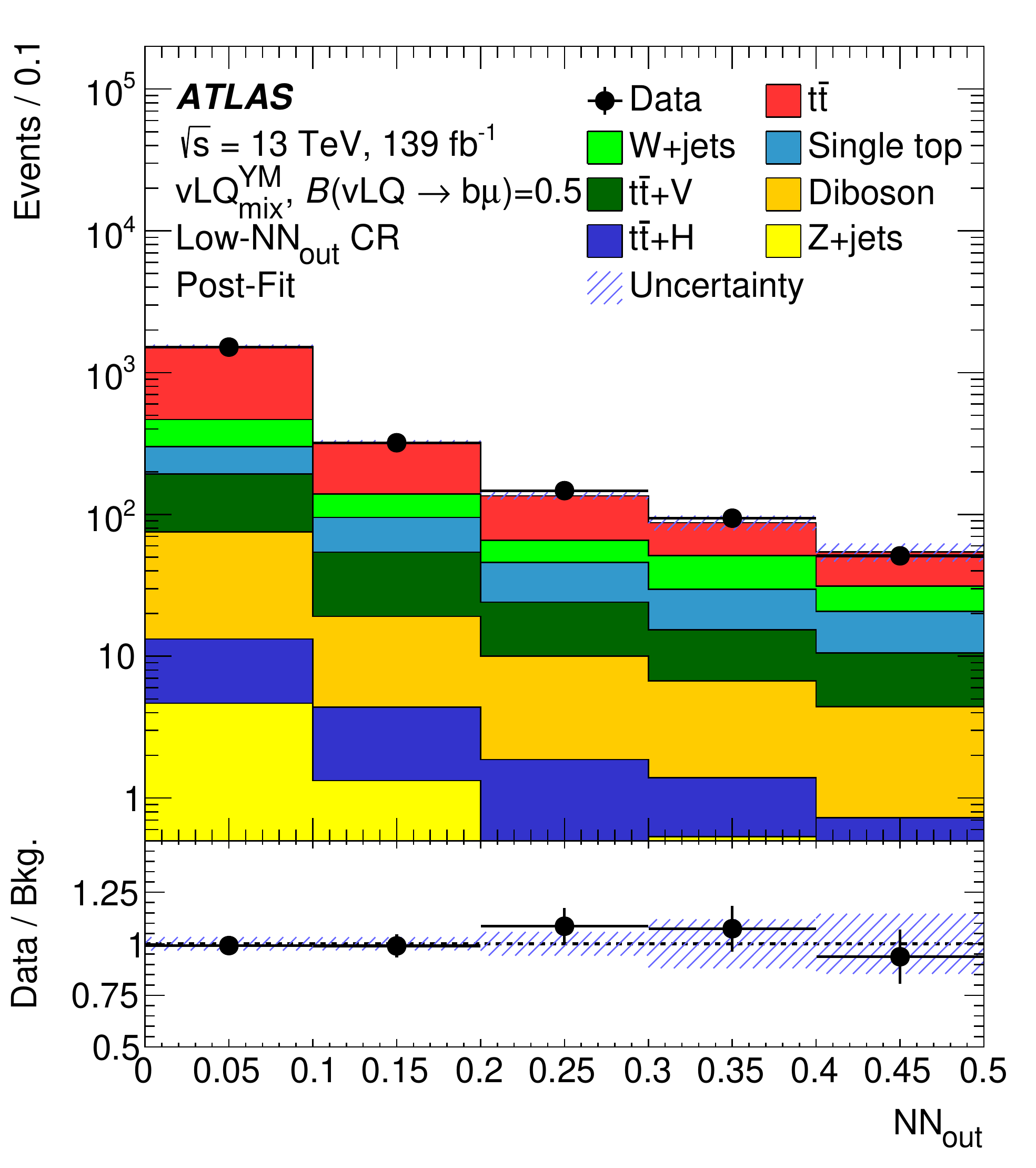}
\caption{}
\label{fig:results:validation:lowNNoutCR}
\end{subfigure}
\caption{Data and background expectation in \subref{fig:results:validation:wjetsCR} the \Wjets CR, \subref{fig:results:validation:singletopCR} the single-top CR, and \subref{fig:results:validation:lowNNoutCR} the \lownnout CR after the simultaneous background-only fit for a training with $\vLQYM \rightarrow b\mu/t\nu$ and $\br=0.5$. The hatched band indicates the total post-fit uncertainty. The ratios of data to background expectation are shown in the bottom panels.}
\label{fig:results:validation}
\end{figure}

The \nnout distributions after the background-only fit are validated with data--MC comparisons in the control regions, as shown in~\Cref{fig:results:validation} for one particular NN training. In general, good agreement is found for all trainings in all control regions, although the
single-top CR is typically more problematic due to the interference effects mentioned above~\cite{ATL-PHYS-PUB-2021-042}.
NPs corresponding to the systematic uncertainties covering the observed differences between data and MC simulation are not constrained significantly because only the overall number of events in each CR enters the fit.
The systematic uncertainties are therefore fully propagated to the SR.
 
\begin{figure}
\centering
\begin{subfigure}[b]{0.33\textwidth}
\centering
\includegraphics[width=\textwidth]{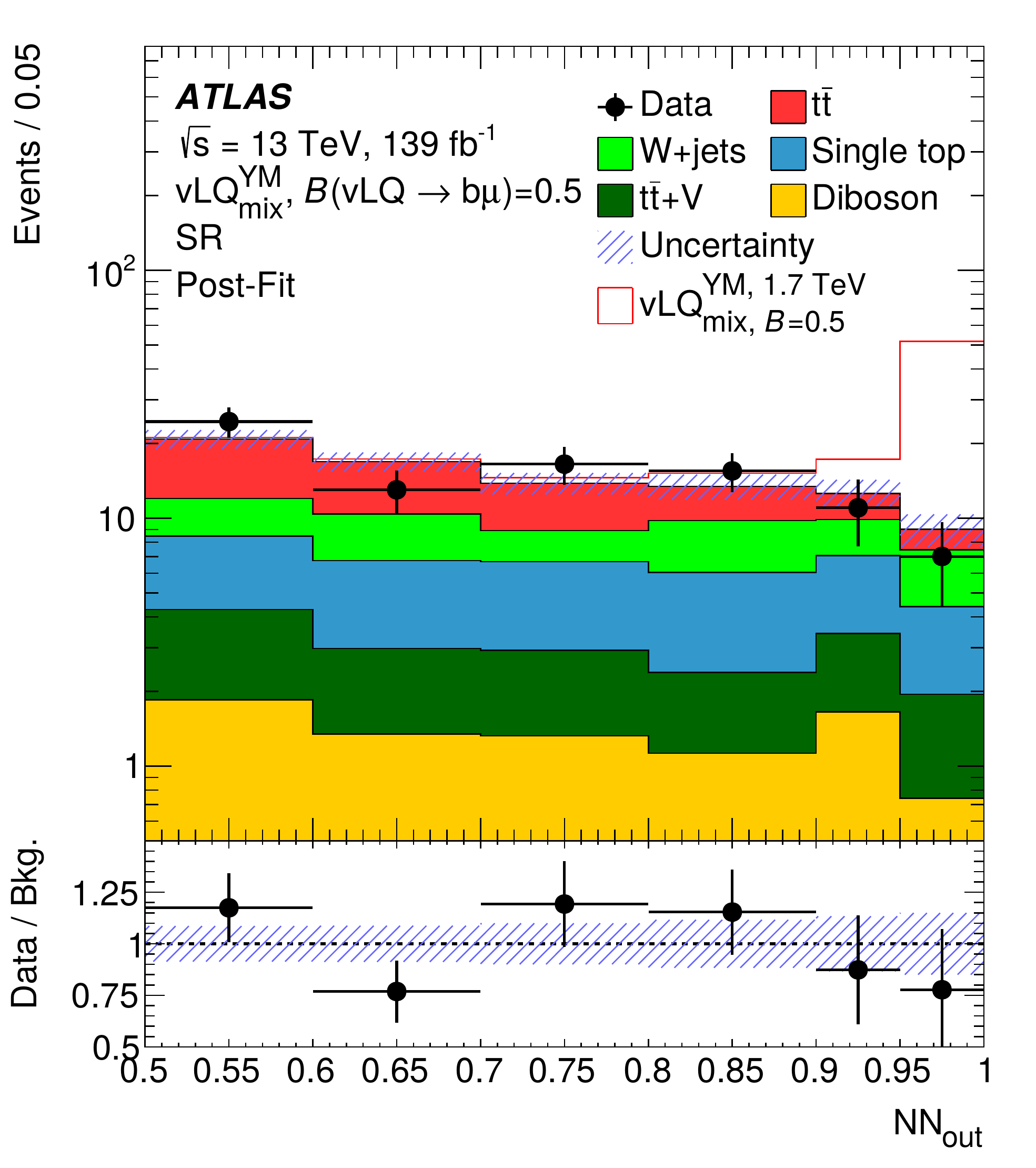}
\caption{}
\label{fig:results:postFit:SR12}
\end{subfigure}\begin{subfigure}[b]{0.33\textwidth}
\centering
\includegraphics[width=\textwidth]{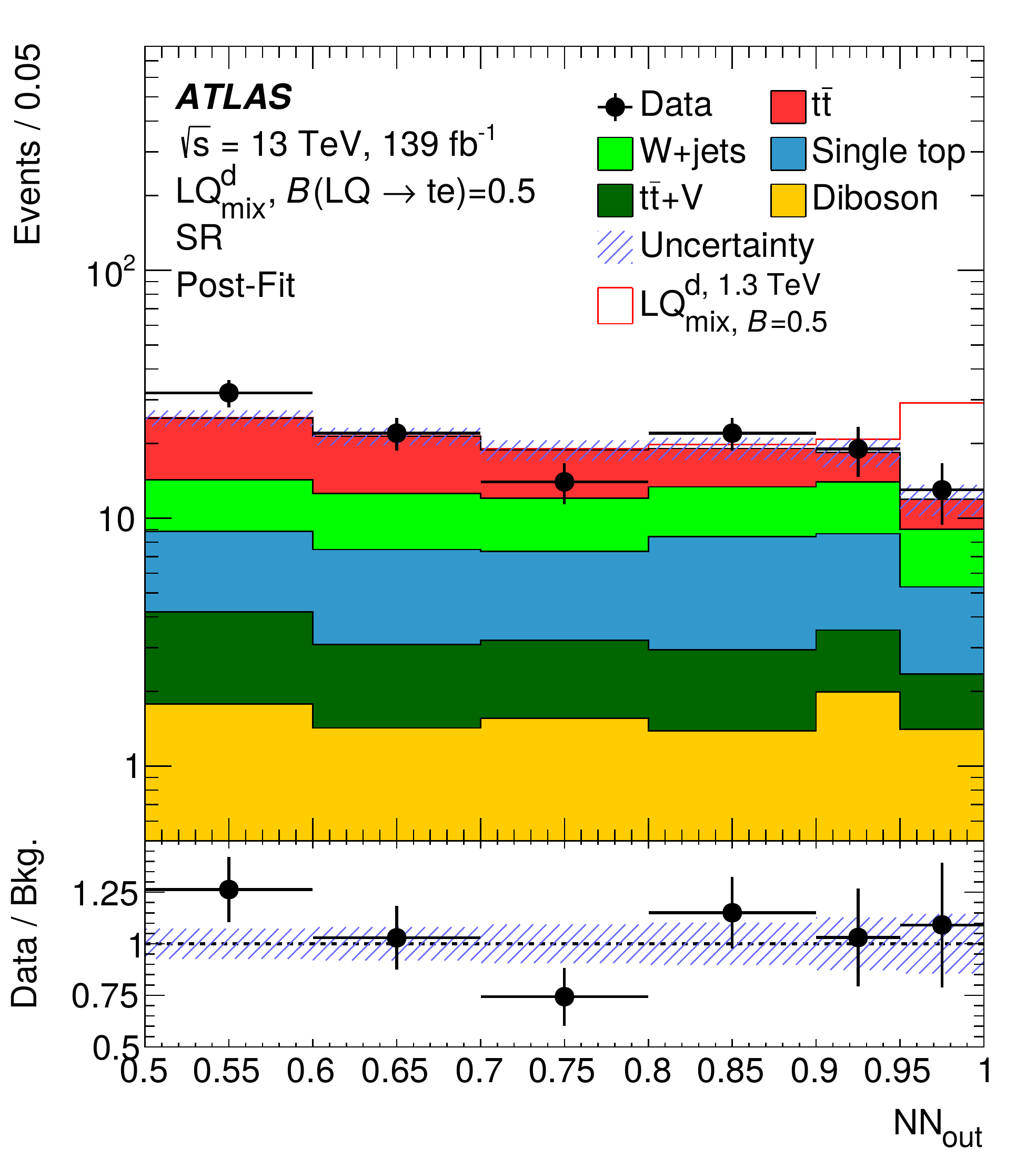}
\caption{}
\label{fig:results:postFit:SR9}
\end{subfigure}\begin{subfigure}[b]{0.33\textwidth}
\centering
\includegraphics[width=\textwidth]{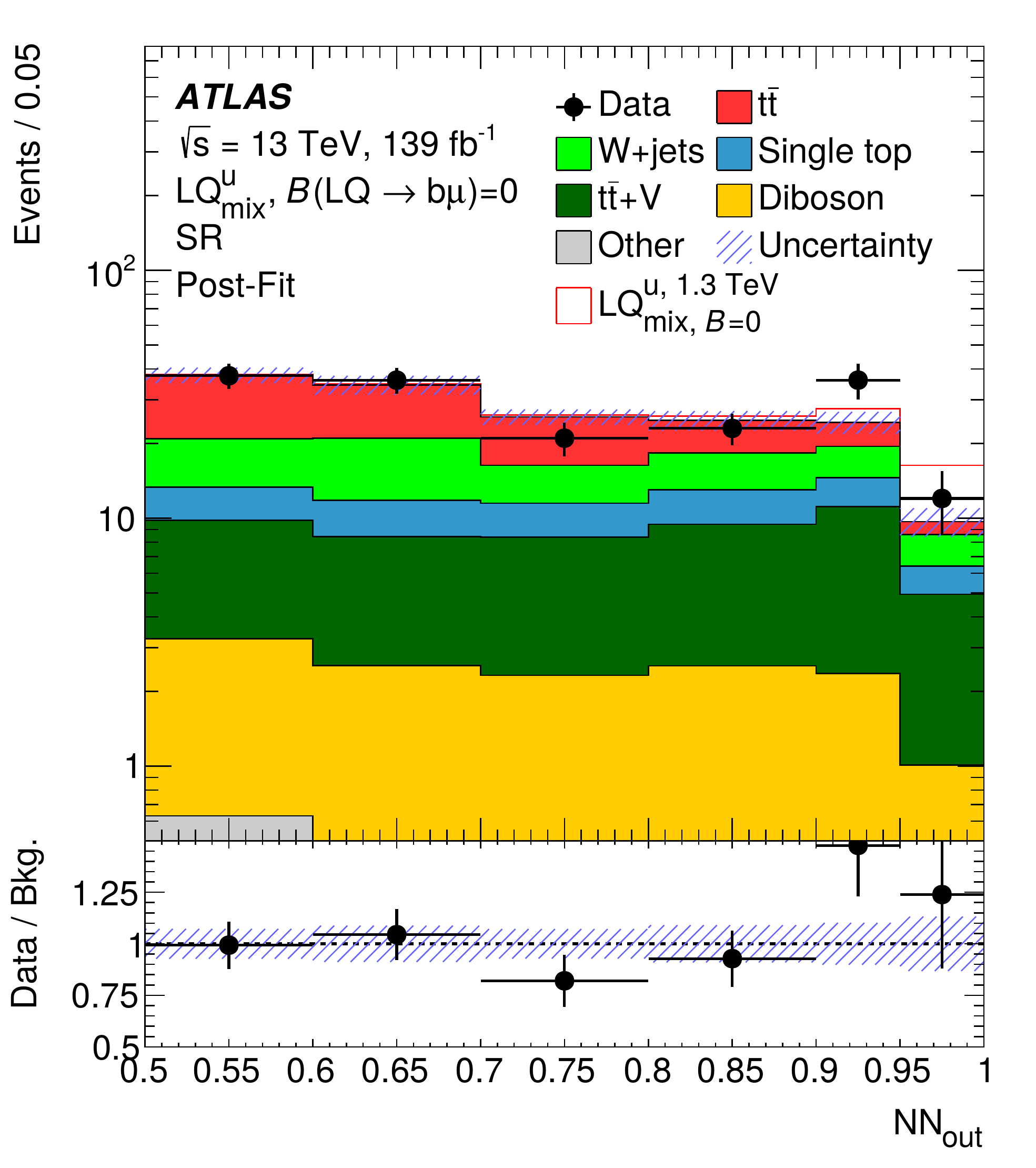}
\caption{}
\label{fig:results:postFit:SR0}
\end{subfigure}
\caption{Data and background expectation in the signal region after the simultaneous background-only fit to data for \subref{fig:results:postFit:SR12} a training with $\vLQYM \rightarrow b\mu/t\nu$ and $\br=0.5$, \subref{fig:results:postFit:SR9} a training with $\lqmixd \rightarrow te/b\nu$ with $\br=0.5$, and \subref{fig:results:postFit:SR0} a training with $\lqmixu \rightarrow t\nu$, i.e.\ $\br=0.0$. Minor background contributions from \ttH and \Zjets are combined into `other'. Expected pre-fit signal distributions with \br corresponding to the respective training are added on top of the background expectation, using a mass of \SI{1700}{\GeV} for vector LQs and \SI{1300}{\GeV} for scalar LQs. The hatched band indicates the total post-fit uncertainty. The ratios of data to background expectation are shown in the bottom panels.}
\label{fig:results:postfit}
\end{figure}

A comparison between data and background expectation in the signal region is shown in~\Cref{fig:results:postfit}
after the background-only fit for three different NN trainings. Good agreement is found for all trainings. The largest
discrepancies at high values of \nnout are observed for the $\lqmixu$ model with $\br=0.0$, i.e.\ for the decay into top-quarks and neutrinos as shown in~\Cref{fig:results:postFit:SR0}.
 
\begin{figure}
\centering
\begin{subfigure}[b]{0.5\textwidth}
\centering
\includegraphics[width=\textwidth]{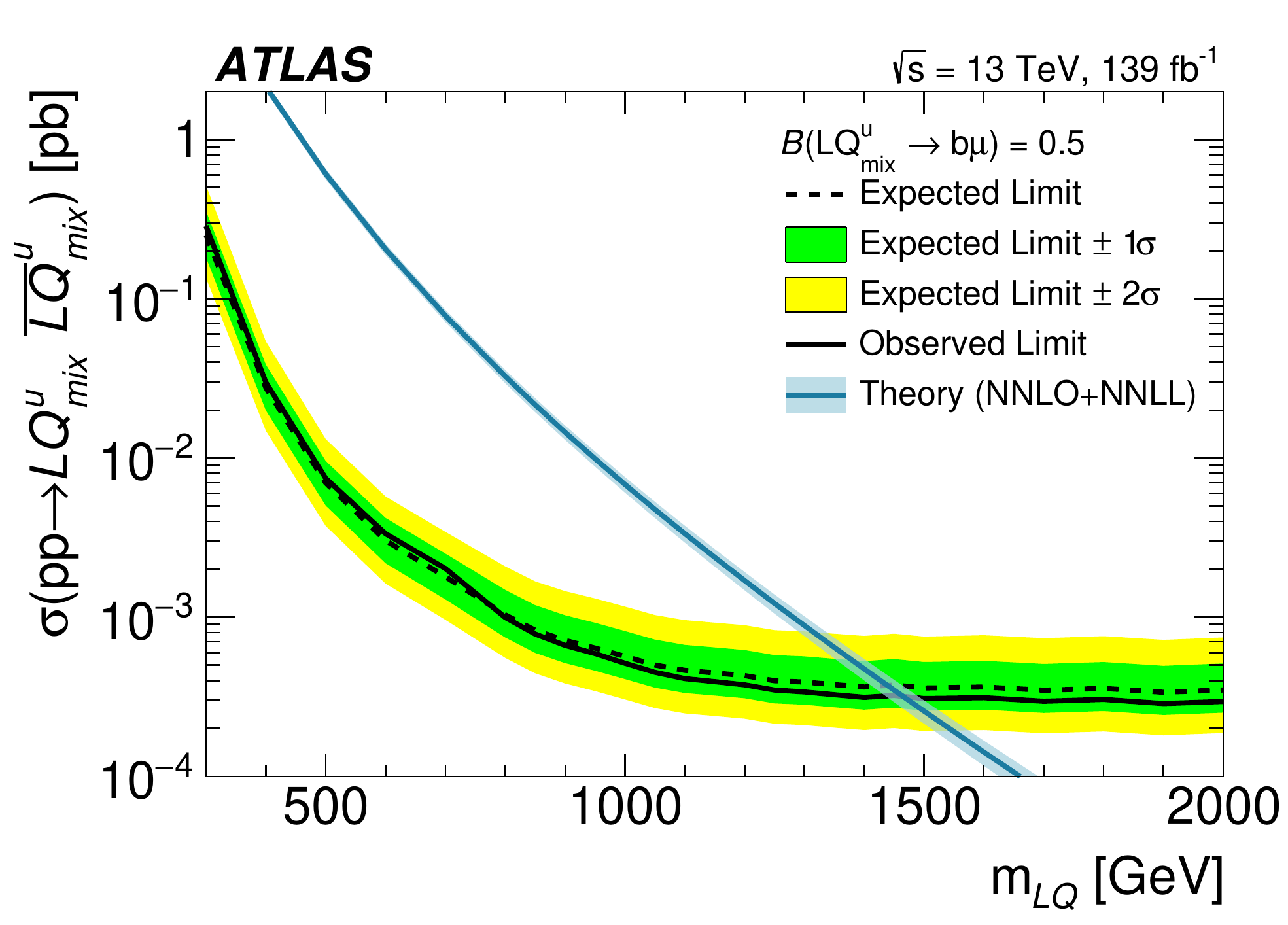}
\caption{}
\label{fig:results:limits1dscalar:up_mu}
\end{subfigure}\begin{subfigure}[b]{0.5\textwidth}
\centering
\includegraphics[width=\textwidth]{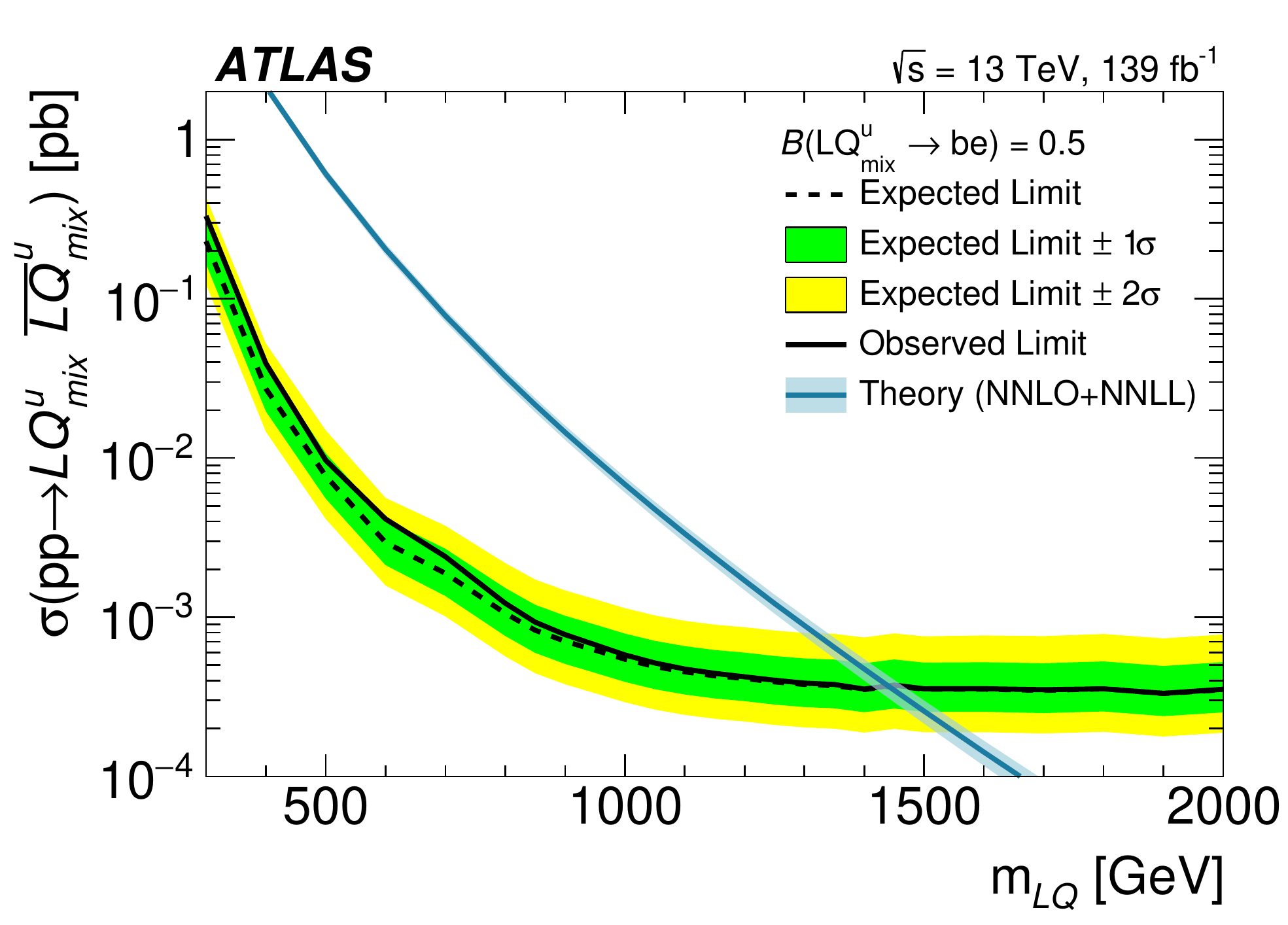}
\caption{}
\label{fig:results:limits1dscalar:up_e}
\end{subfigure}
 
\begin{subfigure}[b]{0.5\textwidth}
\centering
\includegraphics[width=\textwidth]{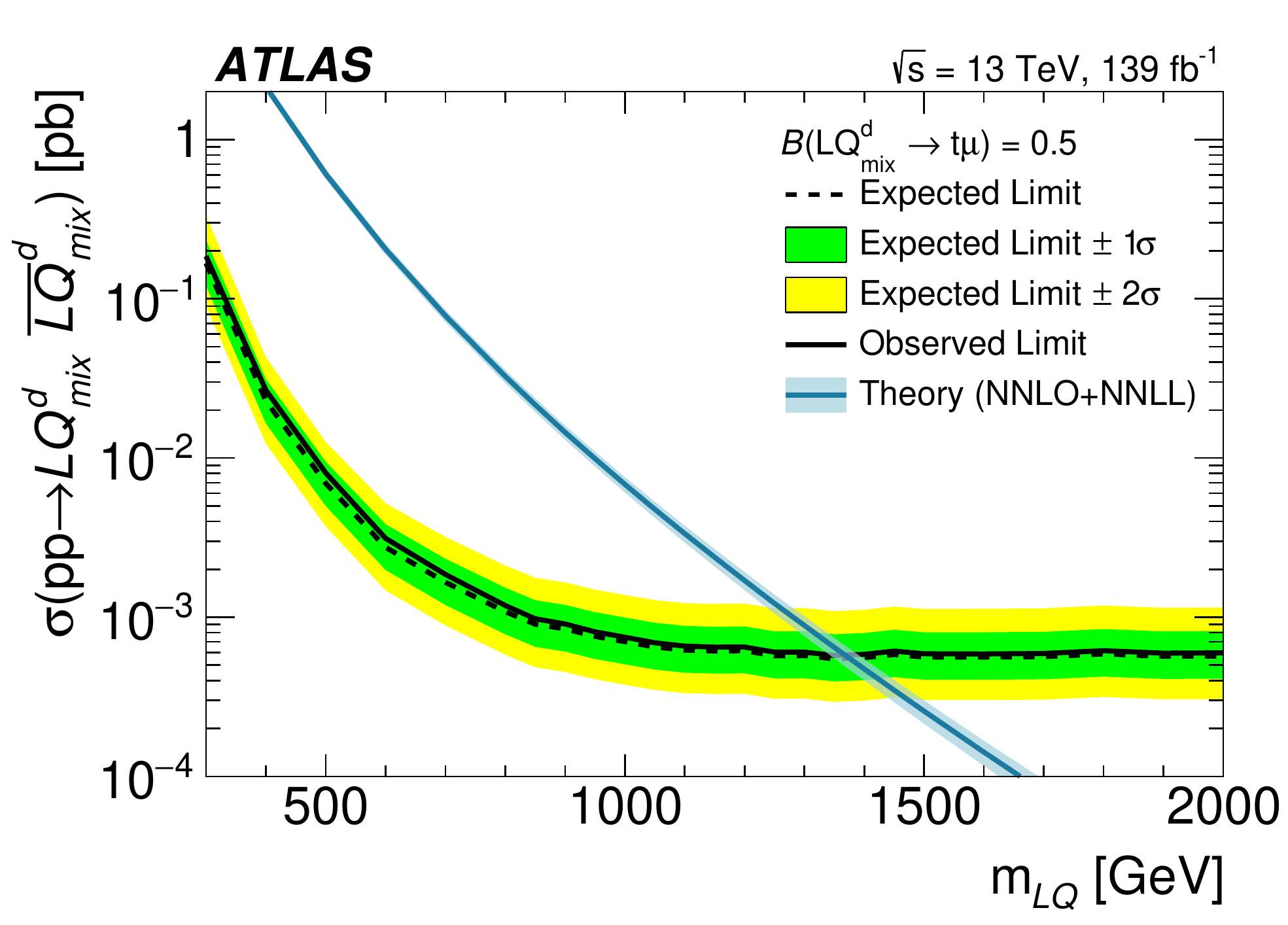}
\caption{}
\label{fig:results:limits1dscalar:down_mu}
\end{subfigure}\begin{subfigure}[b]{0.5\textwidth}
\centering
\includegraphics[width=\textwidth]{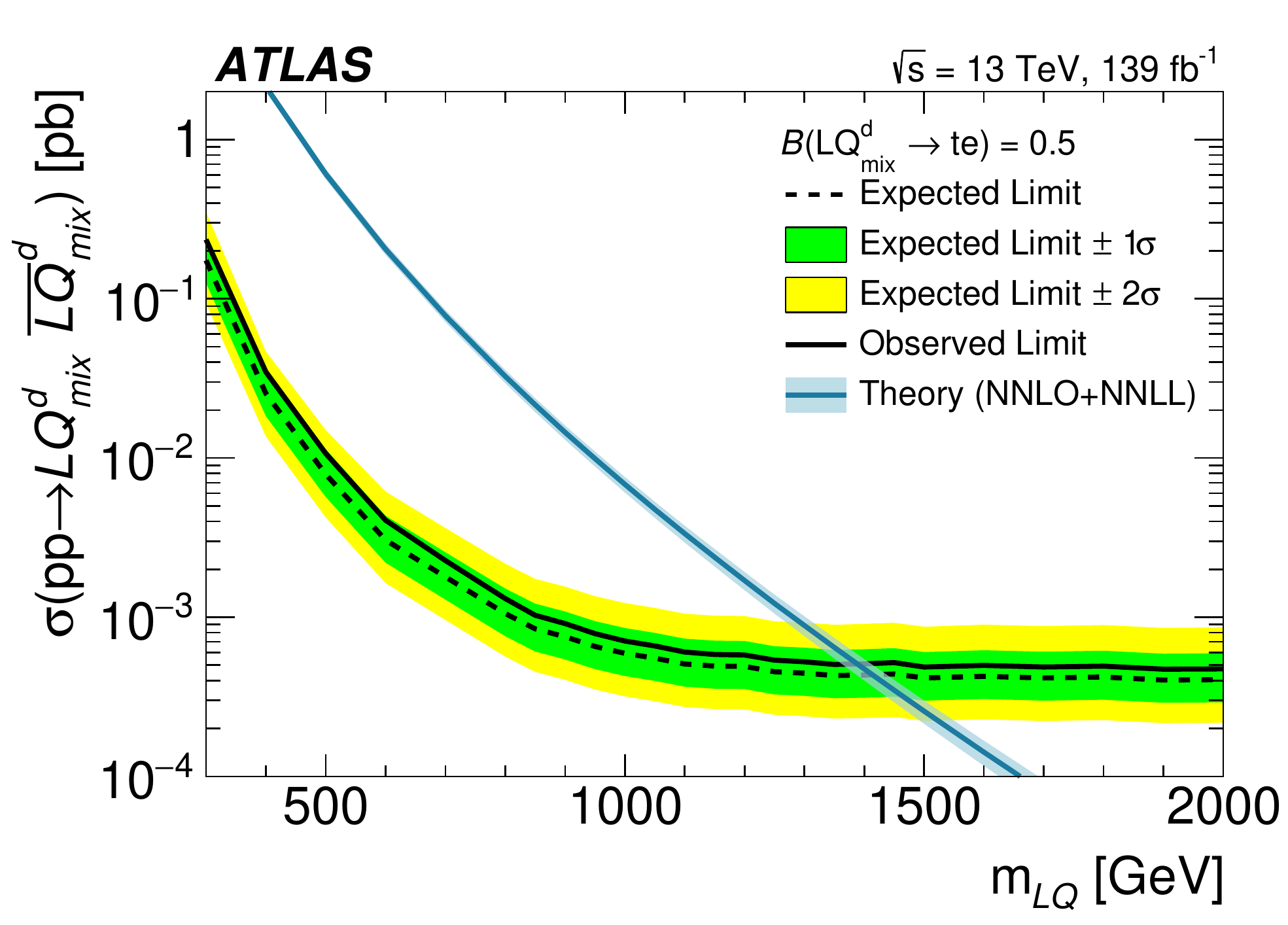}
\caption{}
\label{fig:results:limits1dscalar:down_e}
\end{subfigure}
\caption{Expected (dashed black) and observed (solid black) \SI{95}{\percent} CL upper limits on the cross-section of pair-produced scalar LQs, assuming $\br = 0.5$. The green (yellow) band shows the $\pm1\sigma\,(\pm2\sigma)$ uncertainty region around the expected limit. The theoretical prediction and its $\pm 1\sigma$ uncertainty band are shown in blue. Limits are presented for \subref{fig:results:limits1dscalar:up_mu} up-type scalar LQs decaying into muons, \subref{fig:results:limits1dscalar:up_e} up-type scalar LQs decaying into electrons, \subref{fig:results:limits1dscalar:down_mu} down-type scalar LQs decaying into muons, and \subref{fig:results:limits1dscalar:down_e} down-type scalar LQs decaying into electrons.}
\label{fig:results:limits1dscalar}
\end{figure}

\begin{figure}
\begin{subfigure}[b]{0.5\textwidth}
\centering
\includegraphics[width=\textwidth]{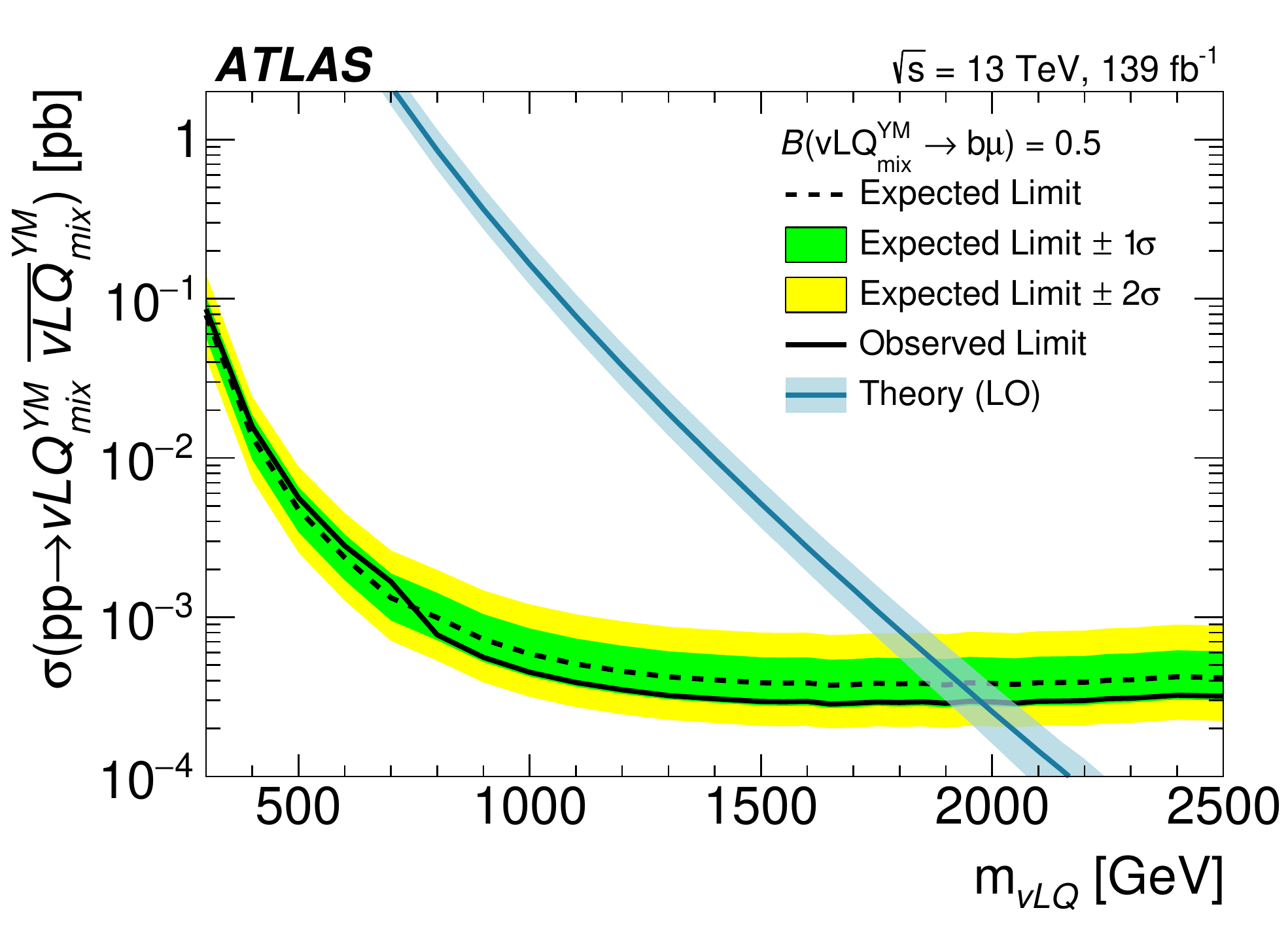}
\caption{}
\label{fig:results:limits1dvector:YM_mu}
\end{subfigure}\begin{subfigure}[b]{0.5\textwidth}
\centering
\includegraphics[width=\textwidth]{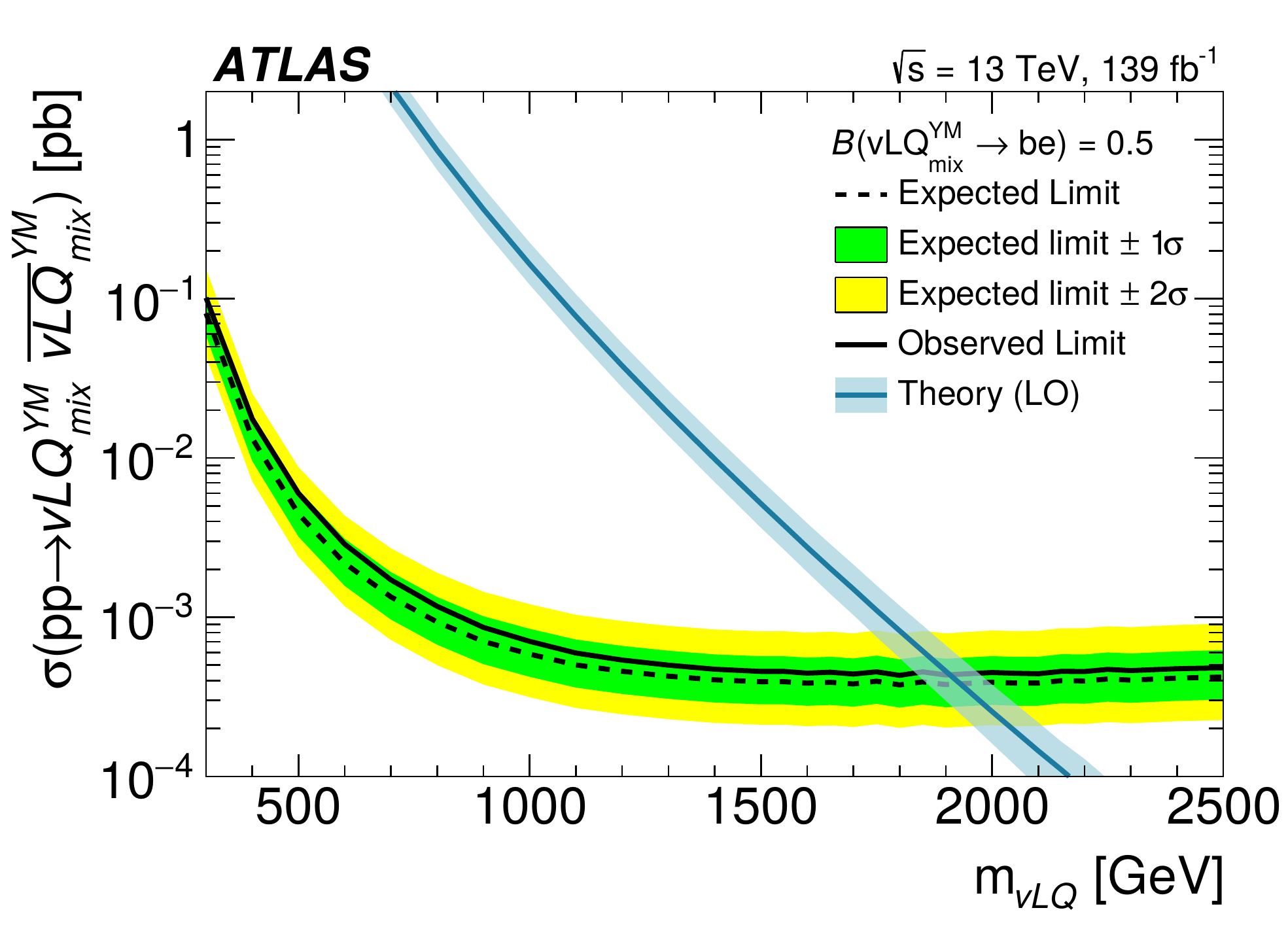}
\caption{}
\label{fig:results:limits1dvector:YM_e}
\end{subfigure}
 
\begin{subfigure}[b]{0.5\textwidth}
\centering
\includegraphics[width=\textwidth]{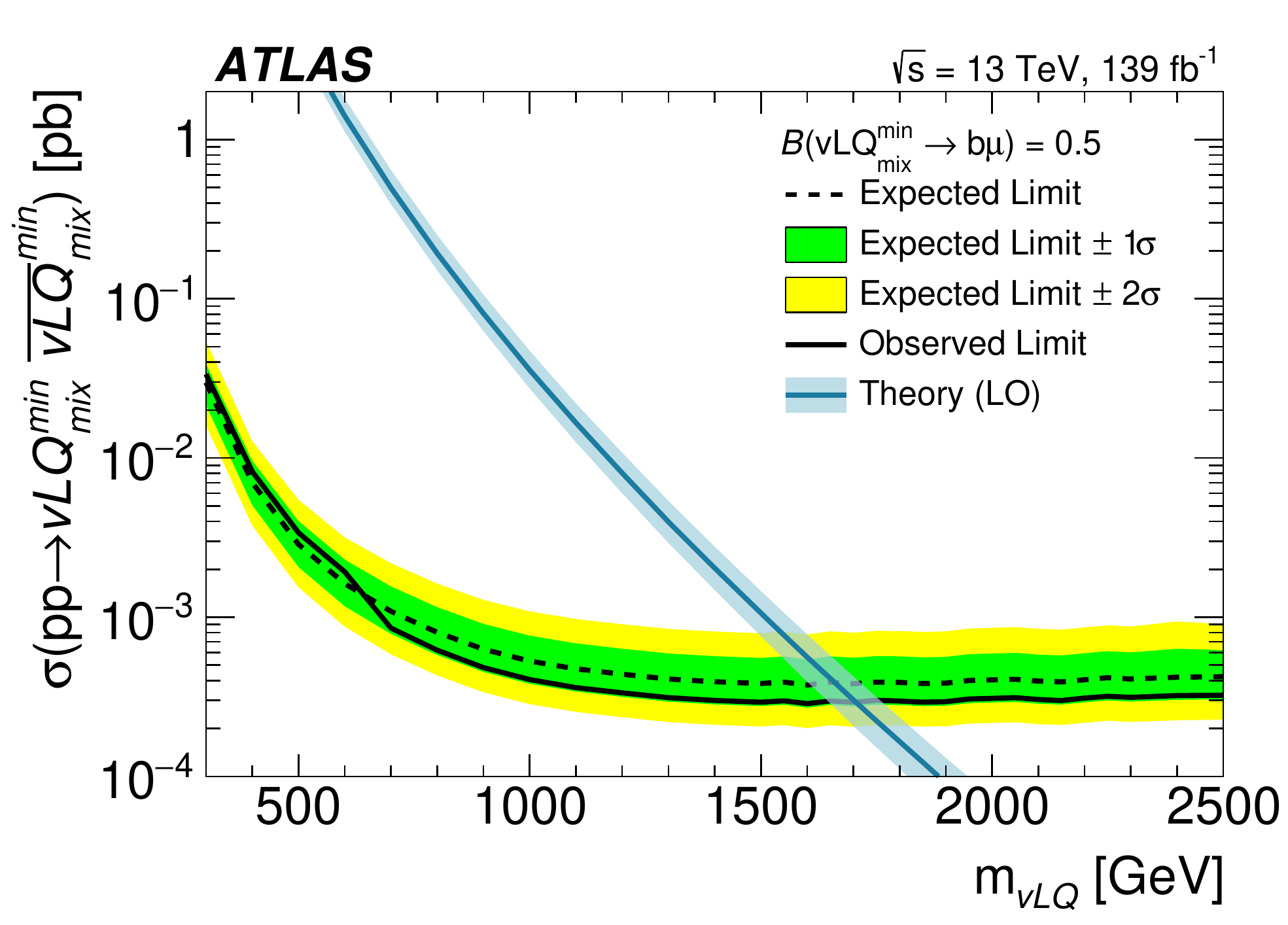}
\caption{}
\label{fig:results:limits1dvector:min_mu}
\end{subfigure}\begin{subfigure}[b]{0.5\textwidth}
\centering
\includegraphics[width=\textwidth]{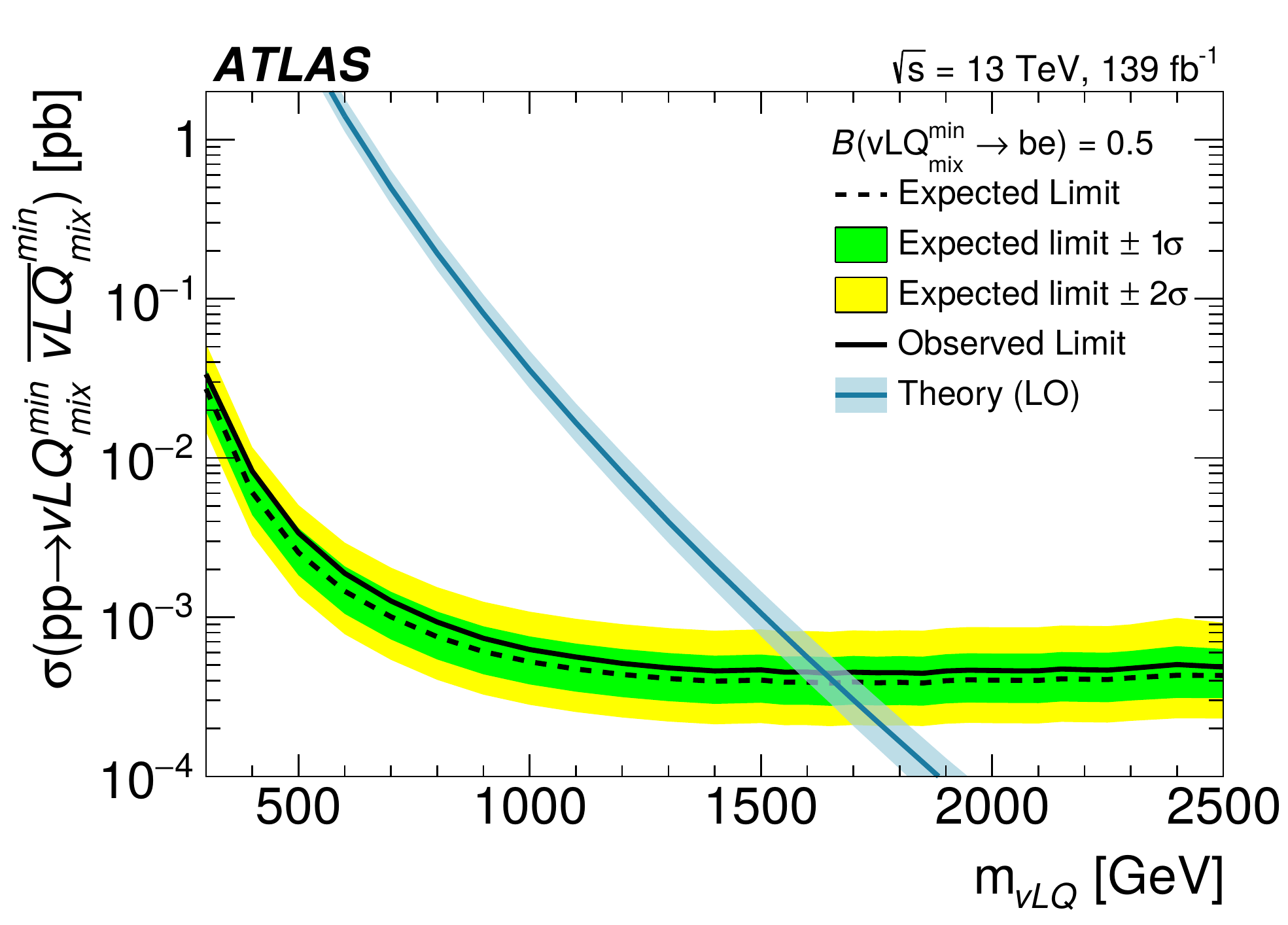}
\caption{}
\label{fig:results:limits1dvector:min_e}
\end{subfigure}
\caption{Expected (dashed black) and observed (solid black) \SI{95}{\percent} CL upper limits on the cross-section of pair-produced vector LQs, assuming $\br = 0.5$. The green (yellow) band shows the $\pm1\sigma\,(\pm2\sigma)$ uncertainty region around the expected limit. The theoretical prediction and its $\pm 1\sigma$ uncertainty band are shown in blue. Limits are presented for \subref{fig:results:limits1dvector:YM_mu} vector LQs in the Yang--Mills coupling scenario decaying into muons, \subref{fig:results:limits1dvector:YM_e} vector LQs in the Yang--Mills coupling scenario decaying into electrons, \subref{fig:results:limits1dvector:min_mu} vector LQs in the minimal coupling scenario decaying into muons, and \subref{fig:results:limits1dvector:min_e} vector LQs in the minimal coupling scenario decaying into electrons.}
\label{fig:results:limits1dvector}
\end{figure}

No significant deviations of the data from the expected SM background are observed.
Upper \SI{95}{\percent} CL limits on the cross-sections of pair-produced LQs can be calculated in
simultaneous signal-plus-background fits to the CRs and the SR, in which the background normalisations and possible
signal contributions are determined.
The largest uncertainty in each of the resulting signal strengths is statistical in nature.
For the three signal hypotheses shown in \Cref{fig:results:postfit}, the statistical uncertainty exceeds \SI{85}{\percent} of the total uncertainty for the two scalar LQ models at $\mlq = \SI{1.3}{\TeV}$ and rises to nearly \SI{100}{\percent} for the vector LQ case at $\mlq = \SI{1.7}{\TeV}$.
The resulting limits on the cross-section for the four scalar LQ models are shown in~\Cref{fig:results:limits1dscalar} as a function of the LQ mass for a fixed $\br = 0.5$.
Corresponding limits for the four vector LQ models are shown in~\Cref{fig:results:limits1dvector}.
 
These cross-section limits are compared with the theoretical cross-section predictions, shown in blue,
resulting in lower limits on the signal mass for $\br = 0.5$.
The uncertainty band around the theory prediction includes PDF, $\alphas$, and renormalisation and factorisation scale uncertainties.
The expected and observed limits for $\br = 0.5$ are summarised in~\Cref{tab:results:limits} for the eight LQ models considered in this analysis.
The total impact of systematic uncertainties on the cross-section limits reaches \SI{15}{\percent} for LQ masses above \SI{1}{\TeV}, corresponding to \SI{20}{\GeV} in the expected mass limit.
 
\begin{table}
\centering
\caption{Expected and observed \SI{95}{\percent} CL lower limits on the LQ mass at $\br = 0.5$ for the eight signal hypotheses considered in this analysis.}
\label{tab:results:limits}
{\def\arraystretch{1.1}
\begin{tabular}{lcc}
\toprule
& Exp.\ limit [\GeV]		& Obs.\ limit [\GeV]	\\
\midrule
$\lqmixu \rightarrow t\nu/b\mu$ 	& $1440_{-60}^{+60}$	& 1460				\\
$\lqmixu \rightarrow t\nu/be$ 		& $1440_{-60}^{+60}$	& 1440				\\
$\lqmixd \rightarrow t\mu/b\nu$ 	& $1380_{-60}^{+50}$	& 1370				\\
$\lqmixd \rightarrow te/b\nu$ 		& $1410_{-60}^{+60}$	& 1390				\\
$\vLQYM \rightarrow t\nu/b\mu$ 		& $1930_{-60}^{+50}$	& 1980				\\
$\vLQYM \rightarrow t\nu/be$ 		& $1930_{-70}^{+50}$	& 1900				\\
$\vLQmin \rightarrow t\nu/b\mu$ 	& $1660_{-50}^{+50}$	& 1710				\\
$\vLQmin \rightarrow t\nu/be$ 		& $1650_{-60}^{+50}$	& 1620				\\
\bottomrule
\end{tabular}
}
\end{table}

\begin{figure}
\centering
\begin{subfigure}[b]{0.5\textwidth}
\centering
\includegraphics[width=\textwidth]{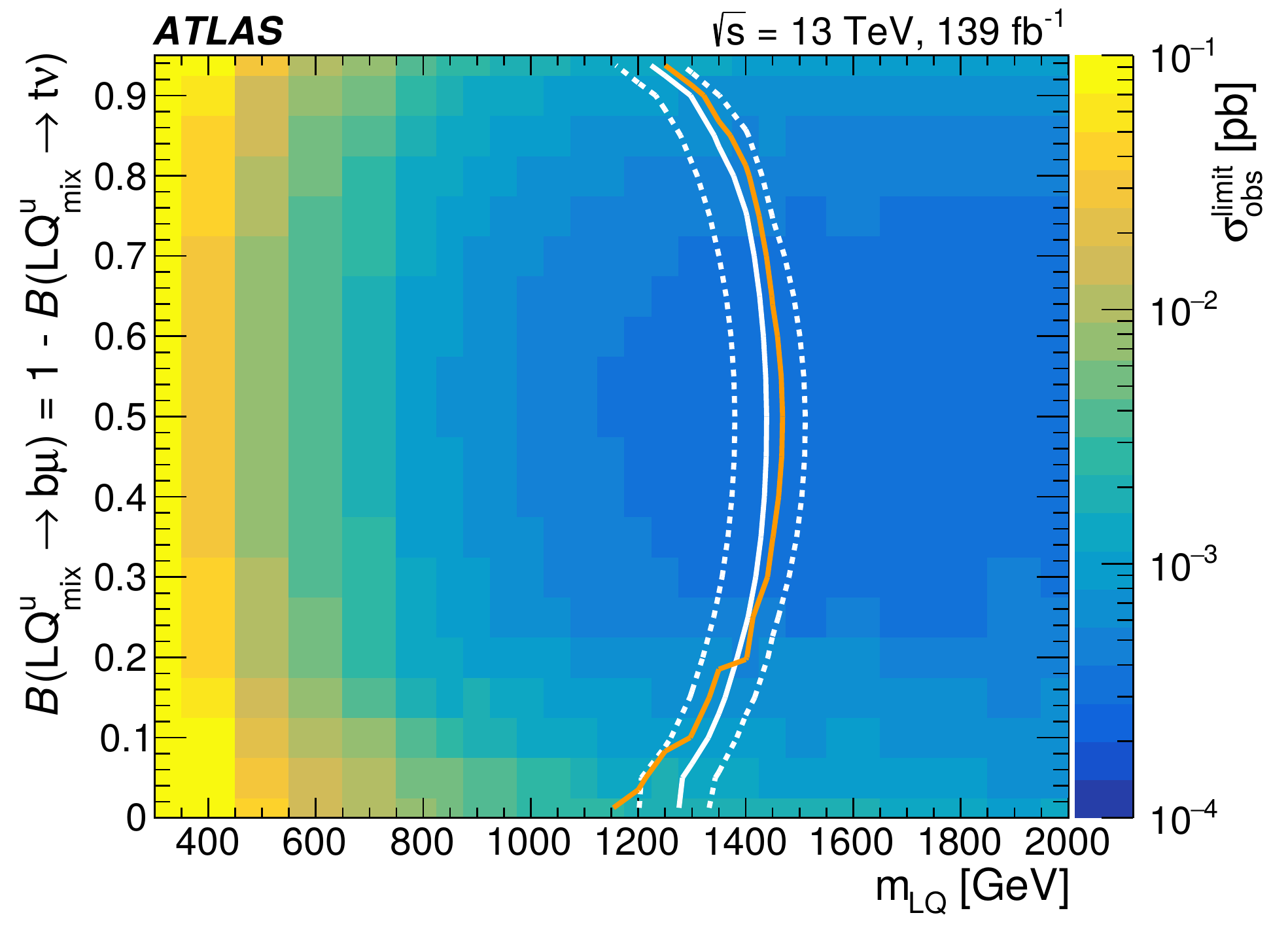}
\caption{}
\label{fig:results:limits2dscalar:up_mu}
\end{subfigure}\begin{subfigure}[b]{0.5\textwidth}
\centering
\includegraphics[width=\textwidth]{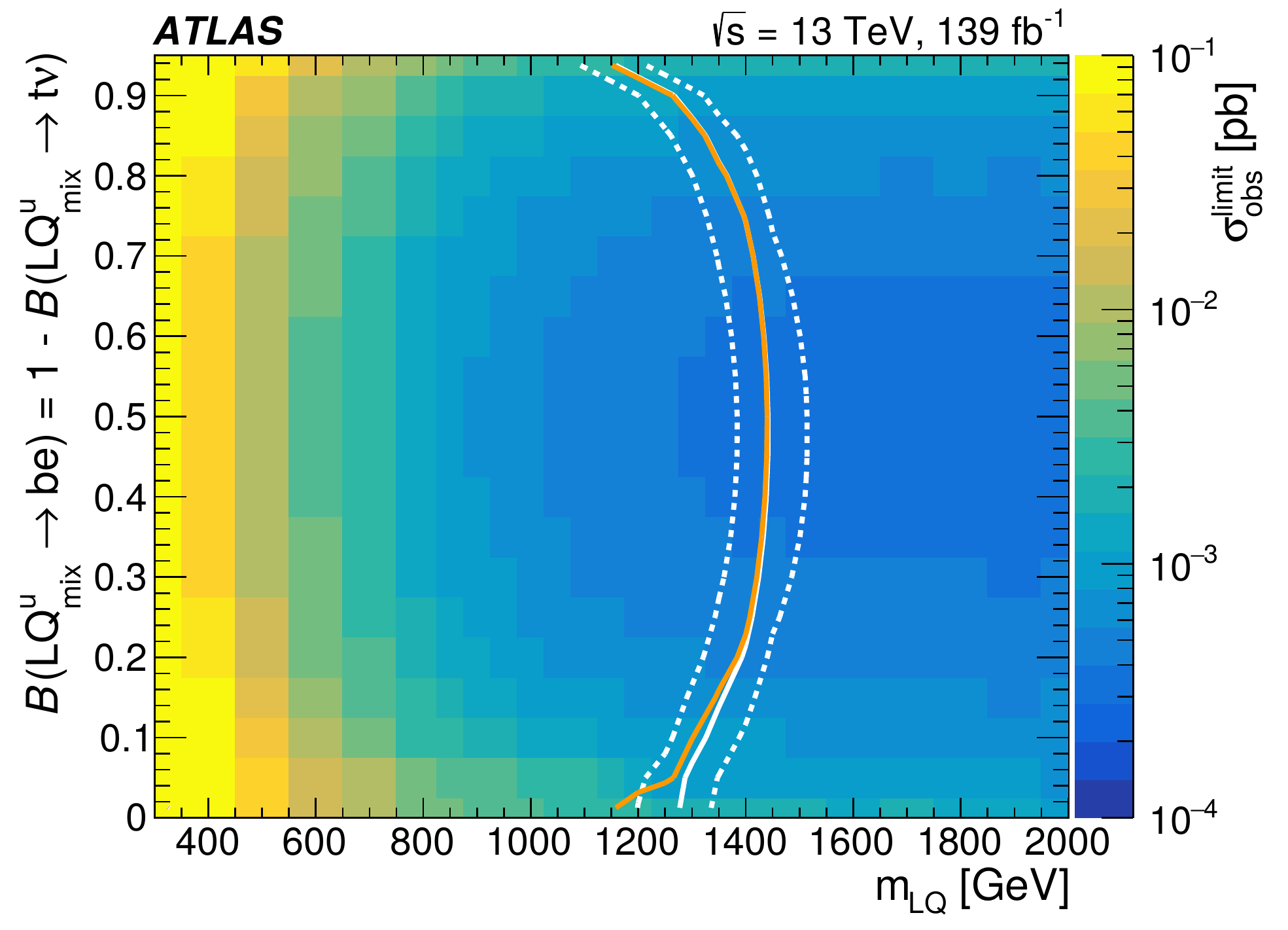}
\caption{}
\label{fig:results:limits2dscalar:up_e}
\end{subfigure}
 
\begin{subfigure}[b]{0.5\textwidth}
\centering
\includegraphics[width=\textwidth]{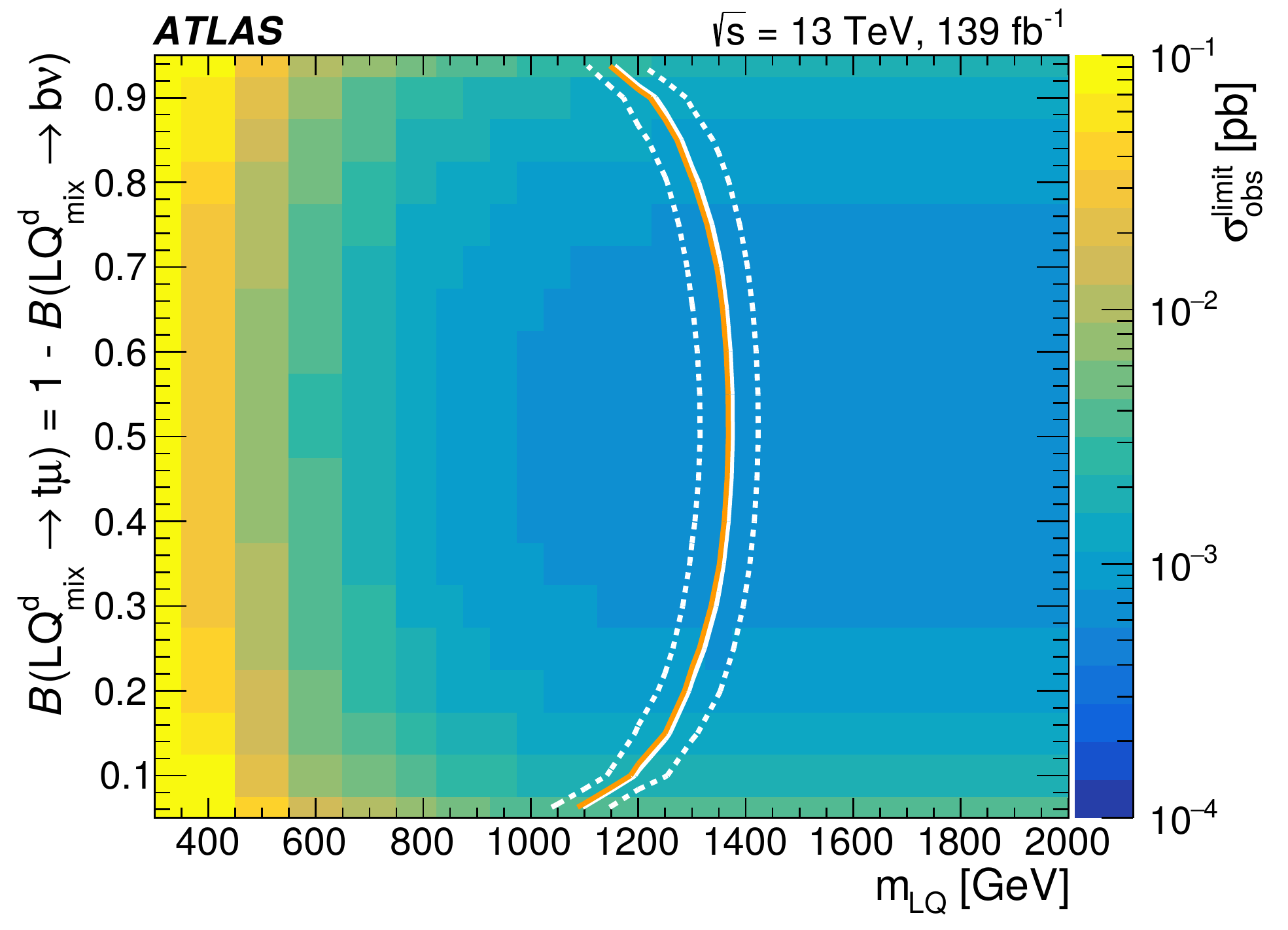}
\caption{}
\label{fig:results:limits2dscalar:down_mu}
\end{subfigure}\begin{subfigure}[b]{0.5\textwidth}
\centering
\includegraphics[width=\textwidth]{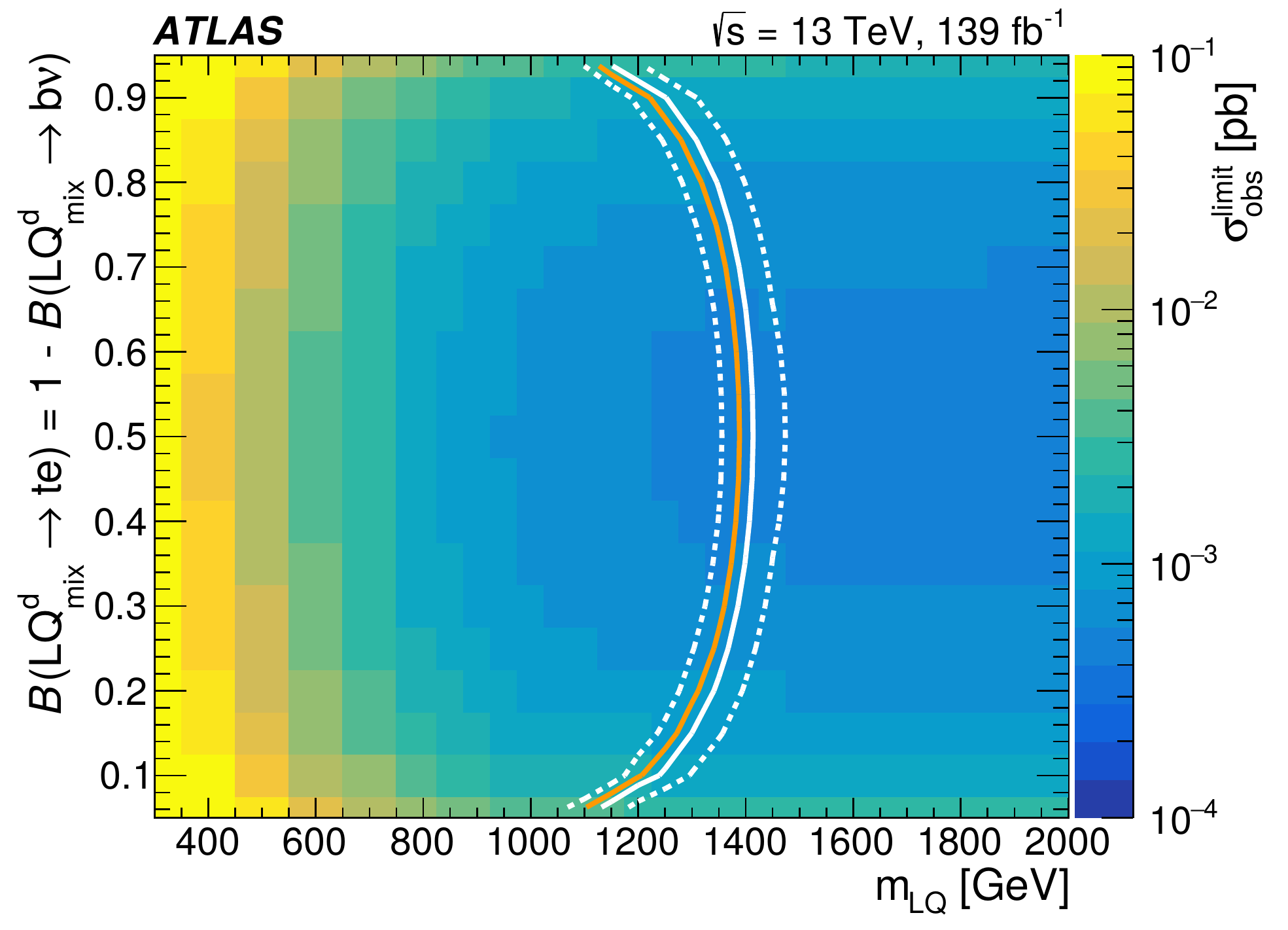}
\caption{}
\label{fig:results:limits2dscalar:down_e}
\end{subfigure}
\caption{
Expected (solid white, $\pm 1\sigma$ ranges dashed) and observed (solid orange) exclusion limits on the leptoquark mass as a function of the branching ratio into charged leptons at \SI{95}{\percent} CL.
The observed upper limit on the signal cross-section in each bin is shown on the z-axis. Limits are presented for \subref{fig:results:limits2dscalar:up_mu} up-type scalar LQs decaying into muons, \subref{fig:results:limits2dscalar:up_e} up-type scalar LQs decaying into electrons, \subref{fig:results:limits2dscalar:down_mu} down-type scalar LQs decaying into muons, and \subref{fig:results:limits2dscalar:down_e} down-type scalar LQs decaying into electrons. For up-type LQs the range in $\br$ is 0--0.95, for down-type it is 0.05--0.95.}
\label{fig:results:limits2dscalar}
\end{figure}

\begin{figure}
\begin{subfigure}[b]{0.5\textwidth}
\centering
\includegraphics[width=\textwidth]{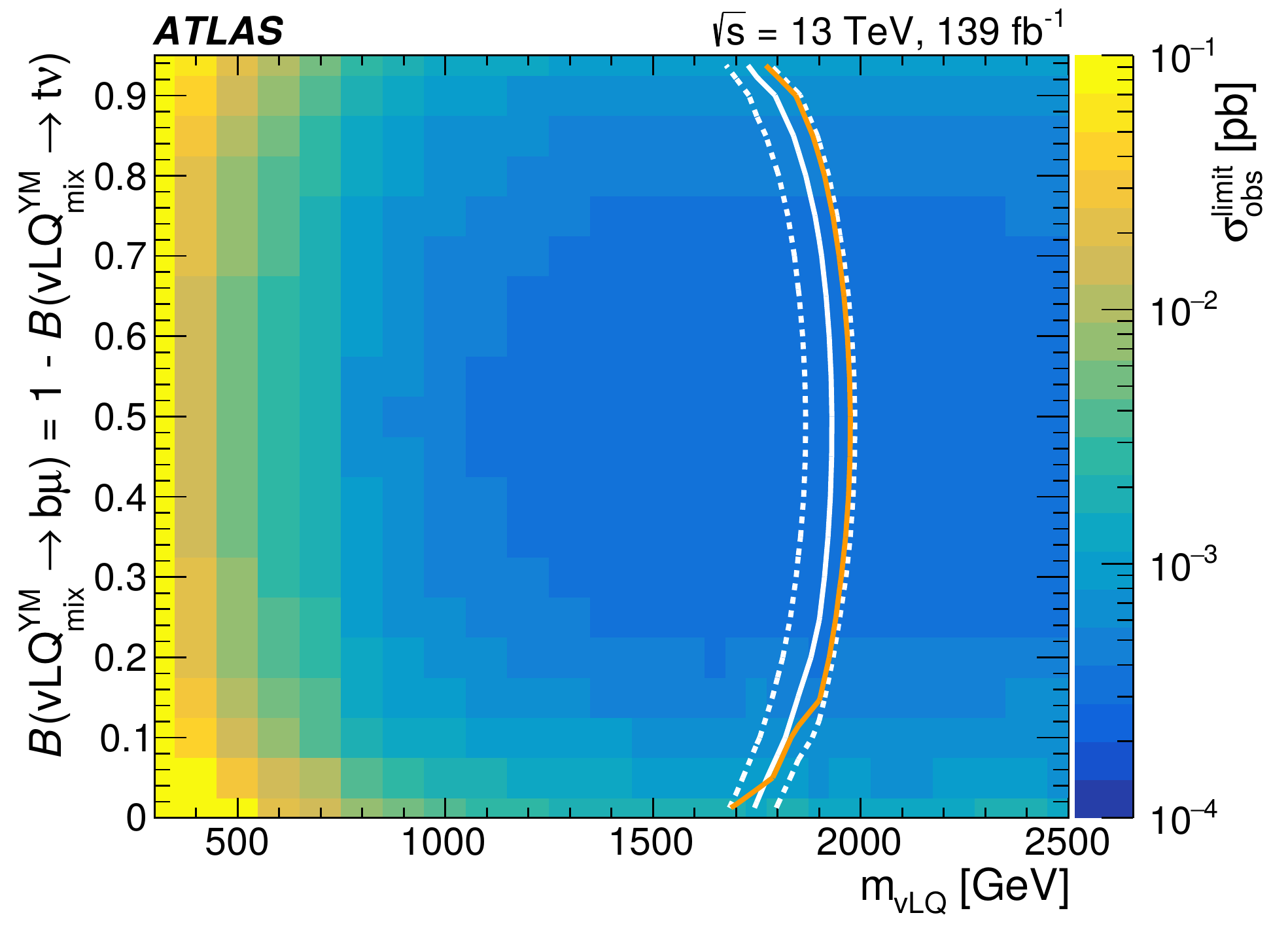}
\caption{}
\label{fig:results:limits2dvector:YM_mu}
\end{subfigure}\begin{subfigure}[b]{0.5\textwidth}
\centering
\includegraphics[width=\textwidth]{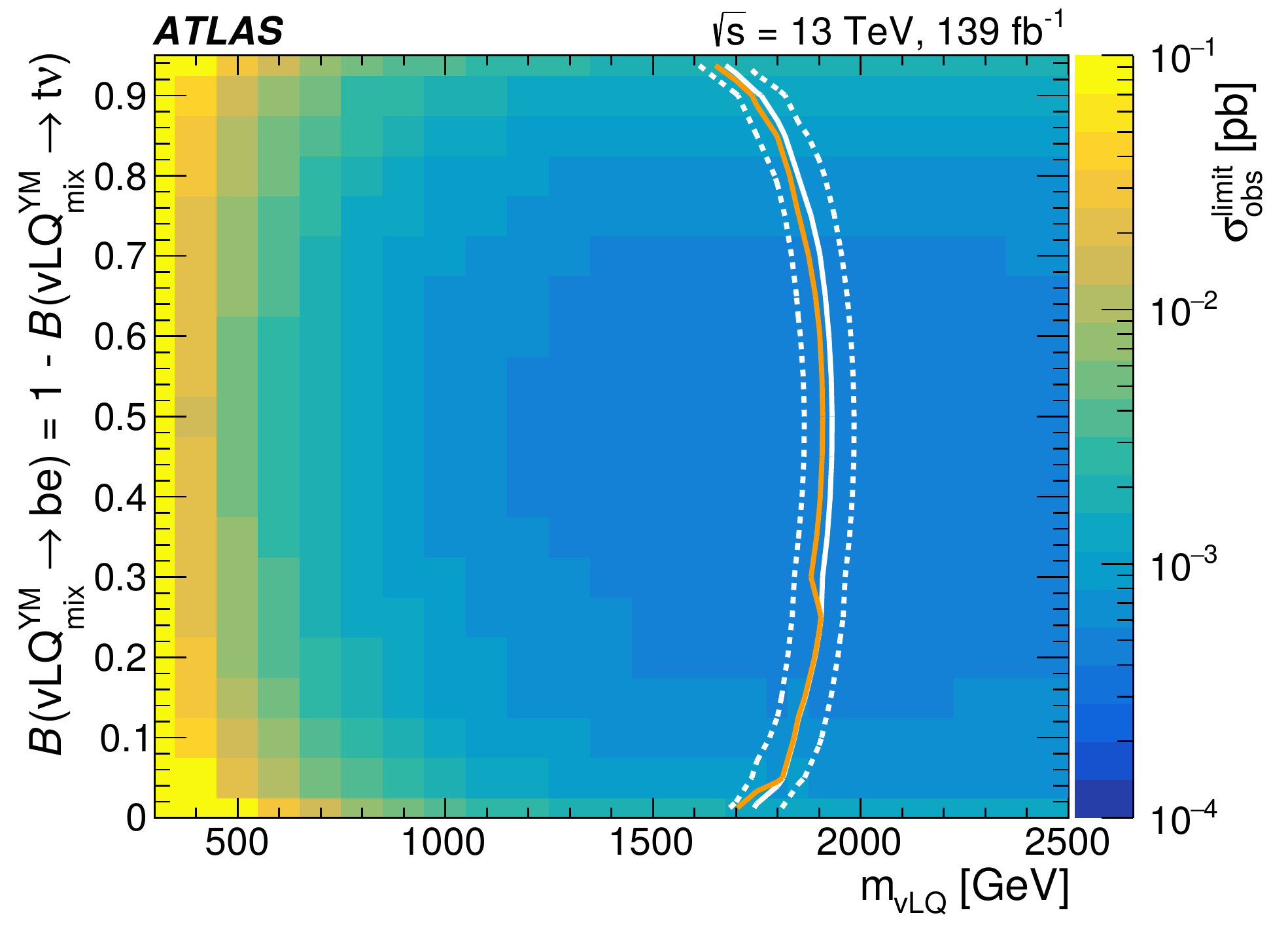}
\caption{}
\label{fig:results:limits2dvector:YM_e}
\end{subfigure}
 
\begin{subfigure}[b]{0.5\textwidth}
\centering
\includegraphics[width=\textwidth]{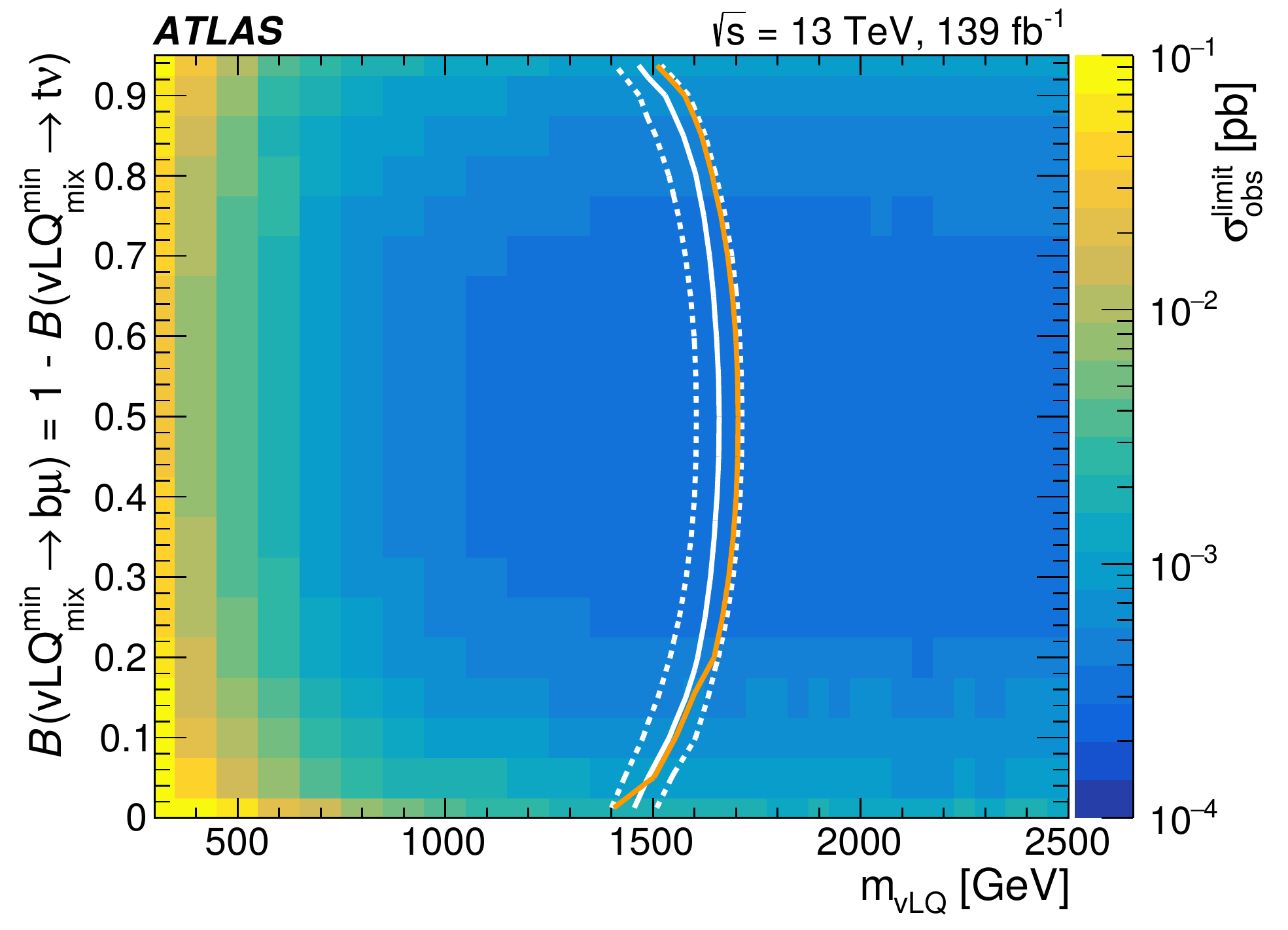}
\caption{}
\label{fig:results:limits2dvector:min_mu}
\end{subfigure}\begin{subfigure}[b]{0.5\textwidth}
\centering
\includegraphics[width=\textwidth]{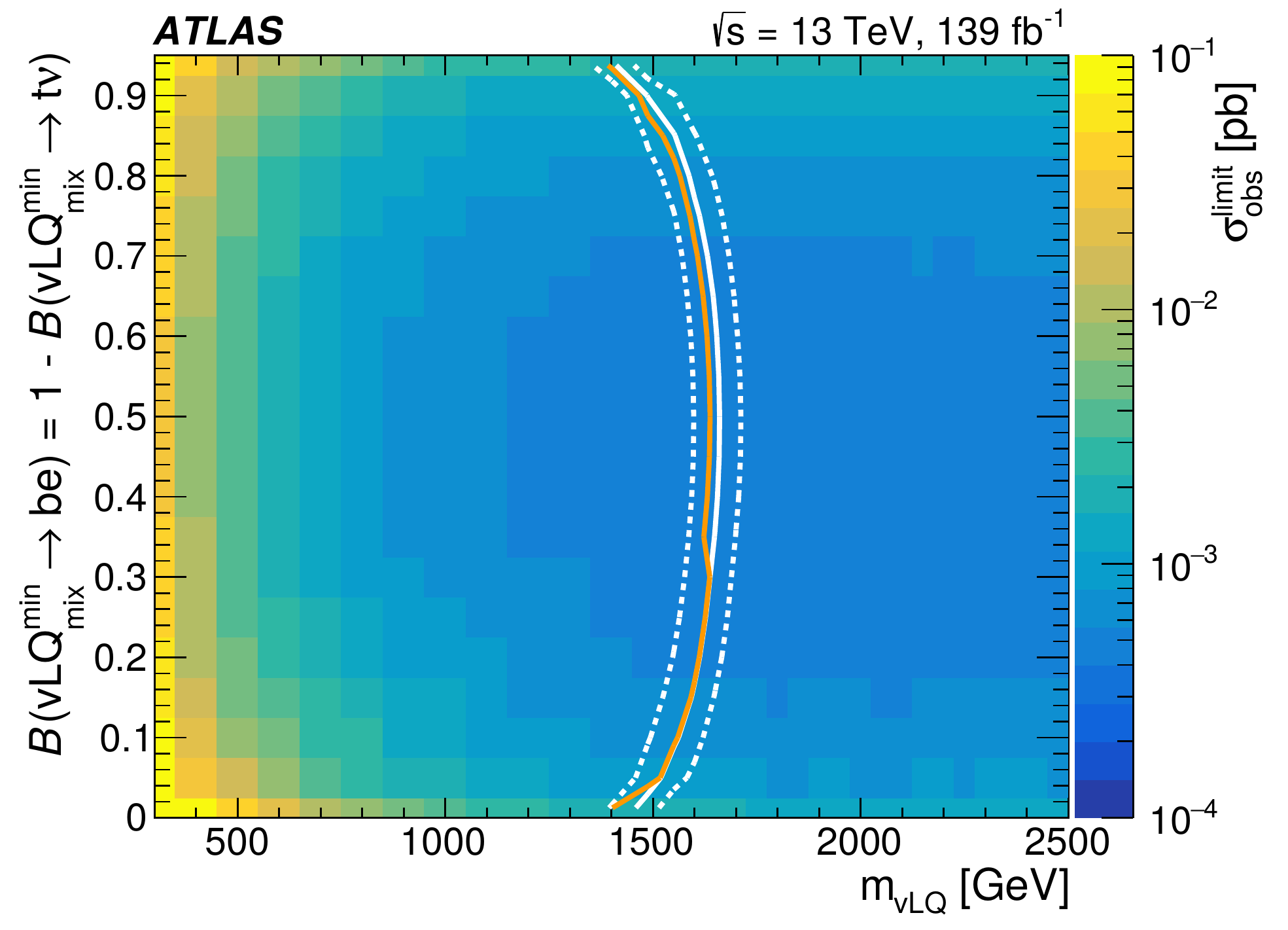}
\caption{}
\label{fig:results:limits2dvector:min_e}
\end{subfigure}
\caption{
Expected (solid white, $\pm 1\sigma$ ranges dashed) and observed (solid orange) exclusion limits on the leptoquark mass as a function of
the branching ratio into charged leptons at \SI{95}{\percent} CL.
The observed upper limit on the signal cross-section in each bin is shown on the z-axis. Limits are presented for \subref{fig:results:limits2dvector:YM_mu} vector LQs in the Yang--Mills coupling scenario decaying into muons, \subref{fig:results:limits2dvector:YM_e} vector LQs in the Yang--Mills coupling scenario decaying into electrons, \subref{fig:results:limits2dvector:min_mu} vector LQs in the minimal coupling scenario decaying into muons, and \subref{fig:results:limits2dvector:min_e} vector LQs in the minimal coupling scenario decaying into electrons.}
\label{fig:results:limits2dvector}
\end{figure}

Limits on LQ pair-production are also evaluated across a wide range of values for the branching ratio of LQs into charged leptons.
For that, the statistical interpretation is performed in steps of 0.05 in $\br$ between 0.0 and 0.95 for up-type scalar and vector LQs and
between 0.05 and 0.95 for down-type scalar LQs.
For up-type LQs, for which NNs have been trained at four different values of $\br$, the NN resulting in the best expected cross-section limit is chosen at each step.
The analysis is not sensitive to final states with zero or two leptons; therefore, $\br = 1.0$ is omitted for all LQs and so
is $\br = 0.0$ for down-type LQs.
The cross-section upper limits and the mass exclusion curves across the \br plane are shown in~\Cref{fig:results:limits2dscalar} for scalar LQs and in~\Cref{fig:results:limits2dvector} for vector LQs.
 
Expected and observed limits on the leptoquark mass as a function of \br agree well everywhere,
except for a small deviation for the $\lqmixu$ model at $\br=0.0$, as already discussed above in the context of~\Cref{fig:results:postfit}.
Differences between up- and down-type LQs can be observed especially for low values of $\br$, where events
with $\LQ \LQ \rightarrow \tlep\nu \thad\nu$ increase the sensitivity. The shapes of the exclusion limits for vector LQs with the Yang--Mills
and the minimal coupling are very similar, since no significant kinematic differences exist at these higher masses.
When comparing the limits for up-type scalar LQs with those for vector LQs, the kinematic differences due to spin correlations in $\LQ \LQ \rightarrow \tlep\nu \thad\nu$ become relevant,
i.e.\ when approaching $\br=0.0$, the expected lower limit on the mass decreases faster for the vector LQ, as expected from~\Cref{fig:analysis:signalregion:leptonpt}.


\FloatBarrier
 
\section{Conclusion}
 
Results of a search for pair-produced scalar and vector leptoquarks decaying into quarks of the third generation and charged or neutral leptons
of the first or second generation are presented, targeting the single-lepton final state.
The analysis is based on data collected by the ATLAS experiment in $\sqrt{s} = \SI{13}{\TeV}$ proton--proton collisions,
corresponding to an integrated luminosity of $\SI{139}{\ifb}$.
Several neural networks are trained for various signal hypotheses, covering a wide range of parameters.
No significant deviations from the Standard Model expectation are observed and upper limits on the production
cross-section are derived for eight models as a function of leptoquark mass and branching ratio into the charged lepton.

In addition, lower limits on the leptoquark mass are set across a range of branching ratios for all models.
At a branching ratio of 0.5 they
reach values of
\SI{1460}{\GeV} (\SI{1440}{\GeV}) for up-type scalar leptoquarks decaying into muons (electrons) and \SI{1370}{\GeV} (\SI{1390}{\GeV}) for down-type scalar leptoquarks decaying into muons (electrons).
For the first time, dedicated neural networks are used to search for $U_1$ vector leptoquarks. At a branching ratio of 0.5 the resulting lower limits on the mass for the decay into muons are \SI{1980}{\GeV} and \SI{1710}{\GeV} for the Yang–Mills and the minimal coupling scenario, respectively.
The decay into electrons is also probed and limits of \SI{1900}{\GeV} (\SI{1620}{\GeV}) for the Yang–Mills (minimal) coupling scenario are derived.

\section*{Acknowledgements}


We thank CERN for the very successful operation of the LHC, as well as the
support staff from our institutions without whom ATLAS could not be
operated efficiently.
 
We acknowledge the support of
ANPCyT, Argentina;
YerPhI, Armenia;
ARC, Australia;
BMWFW and FWF, Austria;
ANAS, Azerbaijan;
SSTC, Belarus;
CNPq and FAPESP, Brazil;
NSERC, NRC and CFI, Canada;
CERN;
ANID, Chile;
CAS, MOST and NSFC, China;
Minciencias, Colombia;
MEYS CR, Czech Republic;
DNRF and DNSRC, Denmark;
IN2P3-CNRS and CEA-DRF/IRFU, France;
SRNSFG, Georgia;
BMBF, HGF and MPG, Germany;
GSRI, Greece;
RGC and Hong Kong SAR, China;
ISF and Benoziyo Center, Israel;
INFN, Italy;
MEXT and JSPS, Japan;
CNRST, Morocco;
NWO, Netherlands;
RCN, Norway;
MEiN, Poland;
FCT, Portugal;
MNE/IFA, Romania;
JINR;
MES of Russia and NRC KI, Russian Federation;
MESTD, Serbia;
MSSR, Slovakia;
ARRS and MIZ\v{S}, Slovenia;
DSI/NRF, South Africa;
MICINN, Spain;
SRC and Wallenberg Foundation, Sweden;
SERI, SNSF and Cantons of Bern and Geneva, Switzerland;
MOST, Taiwan;
TENMAK, T\"urkiye;
STFC, United Kingdom;
DOE and NSF, United States of America.
In addition, individual groups and members have received support from
BCKDF, CANARIE, Compute Canada and CRC, Canada;
COST, ERC, ERDF, Horizon 2020 and Marie Sk{\l}odowska-Curie Actions, European Union;
Investissements d'Avenir Labex, Investissements d'Avenir Idex and ANR, France;
DFG and AvH Foundation, Germany;
Herakleitos, Thales and Aristeia programmes co-financed by EU-ESF and the Greek NSRF, Greece;
BSF-NSF and GIF, Israel;
Norwegian Financial Mechanism 2014-2021, Norway;
NCN and NAWA, Poland;
La Caixa Banking Foundation, CERCA Programme Generalitat de Catalunya and PROMETEO and GenT Programmes Generalitat Valenciana, Spain;
G\"{o}ran Gustafssons Stiftelse, Sweden;
The Royal Society and Leverhulme Trust, United Kingdom.
 
The crucial computing support from all WLCG partners is acknowledged gratefully, in particular from CERN, the ATLAS Tier-1 facilities at TRIUMF (Canada), NDGF (Denmark, Norway, Sweden), CC-IN2P3 (France), KIT/GridKA (Germany), INFN-CNAF (Italy), NL-T1 (Netherlands), PIC (Spain), ASGC (Taiwan), RAL (UK) and BNL (USA), the Tier-2 facilities worldwide and large non-WLCG resource providers. Major contributors of computing resources are listed in Ref.~\cite{ATL-SOFT-PUB-2021-003}.


\clearpage

\printbibliography
 
\clearpage
 
\begin{flushleft}
\hypersetup{urlcolor=black}
{\Large The ATLAS Collaboration}

\bigskip

\AtlasOrcid[0000-0002-6665-4934]{G.~Aad}$^\textrm{\scriptsize 101}$,
\AtlasOrcid[0000-0002-5888-2734]{B.~Abbott}$^\textrm{\scriptsize 119}$,
\AtlasOrcid[0000-0002-7248-3203]{D.C.~Abbott}$^\textrm{\scriptsize 102}$,
\AtlasOrcid[0000-0002-1002-1652]{K.~Abeling}$^\textrm{\scriptsize 55}$,
\AtlasOrcid[0000-0002-8496-9294]{S.H.~Abidi}$^\textrm{\scriptsize 29}$,
\AtlasOrcid[0000-0002-9987-2292]{A.~Aboulhorma}$^\textrm{\scriptsize 35e}$,
\AtlasOrcid[0000-0001-5329-6640]{H.~Abramowicz}$^\textrm{\scriptsize 150}$,
\AtlasOrcid[0000-0002-1599-2896]{H.~Abreu}$^\textrm{\scriptsize 149}$,
\AtlasOrcid[0000-0003-0403-3697]{Y.~Abulaiti}$^\textrm{\scriptsize 116}$,
\AtlasOrcid[0000-0003-0762-7204]{A.C.~Abusleme~Hoffman}$^\textrm{\scriptsize 136a}$
% ...
\bigskip
\\

% ...
$^{29}$Physics Department, Brookhaven National Laboratory, Upton NY; United States of America.\\
% ...
$^{35}$$^{(a)}$Facult\'e des Sciences Ain Chock, R\'eseau Universitaire de Physique des Hautes Energies - Universit\'e Hassan II, Casablanca;$^{(b)}$Facult\'{e} des Sciences, Universit\'{e} Ibn-Tofail, K\'{e}nitra;$^{(c)}$Facult\'e des Sciences Semlalia, Universit\'e Cadi Ayyad, LPHEA-Marrakech;$^{(d)}$LPMR, Facult\'e des Sciences, Universit\'e Mohamed Premier, Oujda;$^{(e)}$Facult\'e des sciences, Universit\'e Mohammed V, Rabat;$^{(f)}$Institute of Applied Physics, Mohammed VI Polytechnic University, Ben Guerir; Morocco.\\
% ...
$^{55}$II. Physikalisches Institut, Georg-August-Universit\"{a}t G\"ottingen, G\"ottingen; Germany.\\
% ...
$^{101}$CPPM, Aix-Marseille Universit\'e, CNRS/IN2P3, Marseille; France.\\
$^{102}$Department of Physics, University of Massachusetts, Amherst MA; United States of America.\\
% ...
$^{116}$Department of Physics, New York University, New York NY; United States of America.\\
% ...
$^{119}$Homer L. Dodge Department of Physics and Astronomy, University of Oklahoma, Norman OK; United States of America.\\
% ...
$^{136}$$^{(a)}$Departamento de F\'isica, Pontificia Universidad Cat\'olica de Chile, Santiago;$^{(b)}$Millennium Institute for Subatomic physics at high energy frontier (SAPHIR), Santiago;$^{(c)}$Instituto de Investigaci\'on Multidisciplinario en Ciencia y Tecnolog\'ia, y Departamento de F\'isica, Universidad de La Serena;$^{(d)}$Universidad Andres Bello, Department of Physics, Santiago;$^{(e)}$Instituto de Alta Investigaci\'on, Universidad de Tarapac\'a, Arica;$^{(f)}$Departamento de F\'isica, Universidad T\'ecnica Federico Santa Mar\'ia, Valpara\'iso; Chile.\\
% ...
$^{149}$Department of Physics, Technion, Israel Institute of Technology, Haifa; Israel.\\
$^{150}$Raymond and Beverly Sackler School of Physics and Astronomy, Tel Aviv University, Tel Aviv; Israel.\\
% ...
$^{a}$ Also Affiliated with an institute covered by a cooperation agreement with CERN.\\
$^{b}$ Also at Borough of Manhattan Community College, City University of New York, New York NY; United States of America.\\
$^{c}$ Also at Bruno Kessler Foundation, Trento; Italy.\\
$^{d}$ Also at Center for High Energy Physics, Peking University; China.\\
$^{e}$ Also at Centro Studi e Ricerche Enrico Fermi; Italy.\\
$^{f}$ Also at CERN, Geneva; Switzerland.\\
% ...

\end{flushleft}


\end{document}